%% file: main.tex
\documentclass[fleqn,usenatbib]{mnras}

\usepackage[T1]{fontenc}

\DeclareRobustCommand{\VAN}[3]{#2}
\let\VANthebibliography\thebibliography\def\thebibliography{\DeclareRobustCommand{\VAN}[3]{##3}\VANthebibliography}

\usepackage{amsmath}	% Advanced maths commands
\usepackage{amssymb}	% Extra maths symbols
\usepackage{newtxtext,newtxmath}
\usepackage{graphicx}
\usepackage{rotating}
\usepackage{caption}
\usepackage{subcaption}
\usepackage{algorithmic}
\usepackage{adjustbox}
\usepackage{longtable}
\usepackage{multirow}
\usepackage{array}
\usepackage{pdflscape}
\usepackage{afterpage}
\usepackage{color,soul}
\usepackage{float}

\newcommand{\feh}{{\left[\frac{\mathrm{Fe}}{\mathrm{H}}\right]}}
\newcommand{\Q}{{Q_{\star}^{\prime}}}

\title[$\Q$ using Tidal Synchronization]{Constraints on Tidal
Quality Factor in Kepler Eclipsing Binaries using Tidal Synchronization: A
Frequency-Dependent Approach}

\author[Ruskin Patel]{
Ruskin Patel,$^{1}$\thanks{E-mail: ruskin.patel@utdallas.edu}
Kaloyan Penev,$^{1}$\thanks{E-mail: kaloyan.penev@utdallas.edu}
Joshua Schussler$^{1}$
\\
$^{1}$Department of Physics, University of Texas at Dallas, Richardson, Texas\\
}

\date{Accepted XXX. Received YYY; in original form ZZZ}

\pubyear{2022}

\begin{document}\label{firstpage}
\pagerange{\pageref{firstpage}--\pageref{lstarpage}}

\maketitle

\begin{abstract}
    Tidal dissipation in binary systems is the primary source for
    synchronization and circularization of the objects in the system. The
    efficiency of the dissipation of tidal energy inside stars or planets
    results in significant changes in observed properties of the binary system
    and is often studied empirically using a parameter, commonly known as the
    modified tidal quality factor ($\Q$). Though often assumed constant, in
    general that parameter will depend on the particular tidal wave experiencing
    the dissipation and the properties of the tidally distorted object. In this
    work we study the frequency dependence of $\Q$ for Sun-like stars. We
    parameterize $\Q$ as a saturating power-law in tidal frequency and obtain
    constraints using the stellar rotation period of 70 eclipsing binaries
    observed by Kepler. We use Bayesian analysis to account for the
    uncertainties in the observational data required for tidal evolution. Our
    analysis shows that $\Q$ is well constrained for tidal periods $ > 15$\,
    days, with a value of $\Q\sim10^8$ for periods $> 30$\,days and a slight
    suggested decrease at shorter periods. For tidal periods $< 15$\,days,
    $\Q$ is no longer tightly constrained, allowing for a broad range of
    possible values that overlaps with the constraints obtained using tidal
    circularization in binaries, which point to much more efficient dissipation:
    $\Q\sim10^6$.
\end{abstract}


%%%%%%%%%%%%%%%%%%%%%%%%%%%%%%%%%%%%%%%%%%%%%%%%%%

%%%%%%%%%%%%%%%%% BODY OF PAPER %%%%%%%%%%%%%%%%%%

\input{introduction}

\input{tidal_evolution}

\input{data}

\input{q_formalism}

\input{bayesian_analysis}

\input{constraints}

\input{discussion}

\input{conclusion}

\section*{Acknowledgements}

This research was supported by NASA grant 80NSSC18K1009.

The authors acknowledge the Texas Advanced Computing Center (TACC) at The
University of Texas at Austin for providing HPC resources that have contributed
to the research results reported within this paper. URL:
\url{http://www.tacc.utexas.edu}

%%%%%%%%%%%%%%%%%%%%%%%%%%%%%%%%%%%%%%%%%%%%%%%%%%
\input{data_availability}


\bibliographystyle{mnras}
\bibliography{bibliography}

%%%%%%%%%%%%%%%%%%%%%%%%%%%%%%%%%%%%%%%%%%%%%%%%%%

%%%%%%%%%%%%%%%%% APPENDICES %%%%%%%%%%%%%%%%%%%%%

\appendix

\input{q_tweak}
\bsp% typesetting comment
\label{lstarpage}
\end{document}

%% file: introduction.tex
\section{Introduction}
\label{sec:introduction}

The study of tides is imperative in understanding the observed properties of
binary systems, as the evolution of the spins and orbital parameters in
binaries are often strongly influenced by tides. The discovery of hot-Jupiter
exoplanets \citep{Mayor1995, Marcy1996} has rekindled interest in understanding
the efficiency of tidal dissipation as tides can provide a reasonable
explanation for their observed short periods and close-to-circular orbits
\citep{Rasio1996, Fabrycky2007, Beaug2012, Vick2019}.

Under the influence of tides, orbits can either shrink or expand depending on
the spins of the binary members and on the orbit. In circular orbits for
example, if a given component of the binary spins slower than the orbit, tides
will transfer angular momentum from the orbit to the spin of the object, leading
to orbital decay. If an object is spinning faster than the orbit, tides cause it
to spin down towards synchronous rotation.  Reliable estimation of the
efficiency of tidal dissipation in such systems is important for understanding
either the survival of hot Jupiters \citep{Debes2010, Hamer2019} or the halting
of the inward migration \citep{Lin1996, Jaime2022}, or both. If the secondary
object is a star, tides usually do not threaten the long-term survival of the
system, because typically binary star systems have enough angular momentum to
achieve a
%state of minimum energy, which is classified into three major observational
%properties of the binary system:
state of minimum energy; this may manifest itself in three major observational
properties of the binary system:
\begin{enumerate}
    \item \textbf{Tidal Synchronization} $\rightarrow$ The rotation period of
        both stars is equal to the orbital period.
    \item \textbf{Tidal Circularization}  $\rightarrow$ Eccentricity of the
        orbit is close to zero.
    \item \textbf{Tidal Inclination}  $\rightarrow$ The angular momentum vector
        of the stars align with the angular momentum vector of orbit.
\end{enumerate}

There are many suggested tidal dissipation processes. They can generally be divided
into two classes depending on the physical mechanism responsible for the
dissipation: equilibrium tides \citep{Zahn_89, Goldreich_Keeley_77,
Penev_et_al_09, Ogilvie_2012, duguid2021, Vidal_2020}, where the energy
dissipation is primarily through turbulent viscosity, and dynamical tides
\citep{zahn1975, Goodman_Dickson_98, Rieutord_Valdettaro_10, Ogilvie_13,
Essick_Weinberg_16, fuller2016, Barker_2020, Ma2021OrbitalDO}, where the waves
excited inside the star interact with tidal perturbations and dissipation is
either through turbulent viscosity or wave breaking. Empirically, the efficiency
of dissipation is often parameterized by what is commonly known as the
\textit{tidal quality factor}: $Q$. The inverse of $Q$ is the ratio of the
energy dissipated from a tidal bulge in a single orbit to the tidal energy
stored in the tidal bulge \citep{Goldreich_Soter_66}. Lower values of Q
correspond to high dissipation and vice versa. In practice, parametric studies
of tides use a modified tidal quality factor: $Q^{\prime} = Q/k_2$
\citep{ogilvie_2013}, where $k_2$ is the second order love number. In this
paper, we use $\Q$ to refer to the modified tidal quality factor of a given
star.

The equilibrium theory of tides assumes that the star maintains hydrostatic
equilibrium while perturbed by the tidal forces. The quadrupole tidal field of
the companion deforms the primary, leading to the formation of tidal bulges.
Under the assumption of weak tides, one can further decompose the time
dependence of the tidal potential as a Fourier series, thinking of each term in
the series as driving an independent tidal bulge and contributing to the
dissipation as a simple superposition. Earlier studies assumed that the
perturbations in the star respond to the tidal potential with a constant lag in
either phase or time \citep{zahn1975, Hut1981, Alexander1973, eggleton1998}. In
general however, each tidal wave can experience different dissipation depending
on its tidal frequency:
\begin{equation}
    \Omega_{\text{tide}} = m\Omega_{\text{orbit}} - n\Omega_{\star}
    \label{eq:tidal_frequency}
\end{equation}
where m and n are integers, $\Omega_{\text{orbit}}$ is the orbital frequency and
$\Omega_{\star}$ is the spin frequency of the star. As a result, the lag, which
is inversely related to $\Q$, is a frequency-dependent parameter rather than
a constant, with the exact frequency dependence determined by the dissipation
mechanism.

Even though dynamical tide models do not strictly follow the above picture, the
resulting energy and angular momentum exchange can still be modeled by an
effective $\Q$ with an associated frequency dependence
%\citep[see][for more thorough discussion]{ogilvie_2013}.
\citetext{see \citealp{ogilvie_2013} for more thorough discussion}.

In this study we focus on stars with radiative cores and convective envelopes.
%\citep[and others]{Goodman_Dickson_98, Lecoanet2013} argue that in such stars,
\citetext{\citealp{Goodman_Dickson_98}, \citealp{Lecoanet2013}, and others}
argue that in such stars, tides can excite g-modes at the boundary of the two
zones, which travel inward through the core. If the wave amplitude is high
enough they can break, dissipating energy and depositing angular momentum. The
resulting effective tidal quality factor has a frequency dependence given by $\Q
\propto \Omega_{\text{tide}}^{-2.6}$ \citep{barker_ogilvie_2011}. If the wave
amplitude is not high enough for wave breaking to take place,
\citet{Essick_Weinberg_16} showed that dissipation can still occur through the
excitation of a network of daughter, grand-daughter, etc. modes and $\Q \propto
\Omega_{\text{tide}}^{-2.4}$.

\citet{OgilvieLin2004, Ogilvie_2007, Ogilvie_2009, Rieutord_Valdettaro_10,
Barker_2020} proposed a different dissipation mechanism, driven by damping of
tidally excited inertial waves by turbulent viscosity in the convective zone of
stars resulting in enhanced dissipation compared to equilibrium tides. Since
inertial waves only exist with frequencies $\left| \omega \right| < \left|
2\Omega \right|$, where $\Omega$ is the spin frequency of the star, the
dissipation will only be enhanced for tidal frequencies in this range. The
resulting dissipation is predicted to be highly frequency dependent with a dense
spectrum of peaks. A widely used method for accounting for enhanced dissipation
due to inertial waves is averaging over the dissipation spectrum in the frequency
domain to calculate the net contribution of $\Q$ from dynamical tides
\citep{Ogilvie_13}.

Yet another dissipation mechanism invokes resonance locking if the tidal
frequency is in resonance with a low-frequency inertial wave in the convective
envelope or a standing g-mode in the radiative zone \citep{Witte1999, Witte2001,
fuller2016, Fuller_2017}. This could lead to enhanced dissipation if the
resonance is held for a long period without breaking the lock. Most recently,
\citet{Ma2021OrbitalDO} predict resonance locking can be modeled with
$\Q \propto \Omega_{\text{tide}}^{13/3}$.

Validating and clarifying this broad range of models requires measuring $\Q$
from observations. This is far from the first effort to accomplish this.
Recently, \citet{Bolmont_mathis_2017} presented a tidal evolution model
restricted to circular orbits, with zero spin-orbit misalignment. The authors
use the frequency-averaged dissipation due to inertial waves
\citep{ogilvie_2013} with a constant time lag model \citep{Hut1981,
eggleton1998} to approximate $\Q$ as a constant parameter.
\citet{Benbakoura2019} makes a distinction between the outer convective and
inner radiative zones of the stars to allow for angular momentum exchange
%between the two zones of the stars. The authors also improve on the
between the two zones. The authors also improve on the
\citet{Bolmont_mathis_2017} model by differentiating between $\Q$ obtained from
equilibrium tides and dynamical tides, but the tidal potential is restricted to
only lower-order eccentricity terms with no spin-orbit misalignment.

In this paper we explore the dependence of $\Q$ on tidal frequency using
low-mass eclipsing binaries observed by Kepler, with parameters taken from
\citet{Windemuth_2019}, combined with rotation period measurements of the
primary star by \citet{Lurie_2017}. Unlike the earlier efforts mentioned above,
we do not assume a particular tidal model, but instead assume a generic
parameterization of frequency dependent $\Q$. We use Bayesian analysis to fully
account for observational uncertainties, and a tidal evolution model designed to
analyze any parametric dependence of $\Q$ on binary properties, along with the
effects of higher order eccentric terms in tidal potential, changes in internal
structure of the star, and spin-down of the star due to magnetic winds.

The rest of the paper is as follows. In Section~\ref{sec:tidal_model}, we
provide details of the tidal evolution model used for the analysis.
Section~\ref{sec:input_data} describes the source of the observational data
used. Section~\ref{sec:q_formalism} shows the formalism for the
frequency-dependent $\Q$ model. Section~\ref{sec:bayesian_analysis} describes the
methodology for Bayesian analysis. Our results are presented in
Section~\ref{sec:constraints}. Section~\ref{sec:discussion} discusses the
$\Q$ constraints obtained. We conclude in Section~\ref{sec:conclusion}.

%% file: tidal_evolution.tex
\section{Tidal Evolution Model}
\label{sec:tidal_model}

We use an updated version of the publically available tidal evolution module
\textbf{P}lanerary \textbf{O}rbital \textbf{E}volution due to \textbf{T}ides
(POET hereafter; \citet{Penev_2014} ) to simulate the stellar and orbital
evolution under the effects of tides. POET follows stellar evolution by
interpolating among a grid of evolutionary tracks calculated using MESA
\citep{Paxton_et_al_11}. MESA splits the star into many concentric shells, but
for the purposes of POET, those are combined into two zones to track stellar
evolution: convective envelope and radiative core. The two zones are assumed to
have solid body rotation but are allowed to spin differently. They are also
assumed to converge toward synchronous rotation on what is referred to as the
core-envelope coupling timescale \citep{Irwin_et_al_07,Gallet_Bouvier_2013}, as
well as exchange mass (which carries with it its specific angular momentum).
The core-envelope coupling torques are calculated as:

\begin{equation}
    \mathbf{T}_\mathrm{conv,c-e}
    =
    -\mathbf{T}_\mathrm{rad,c-e}
    =
    \frac{
        I_\mathrm{conv}\mathbf{L}_\mathrm{rad} -
        I_\mathrm{rad}\mathbf{L}_\mathrm{conv}
    }{
        (I_\mathrm{conv}+I_\mathrm{rad})\tau_\mathrm{c-e}
    }
    -
    \frac{2L_\mathrm{conv}}{3I_\mathrm{conv}} R_\mathrm{rad}^2
    \dot{M}_\mathrm{rad}
\end{equation}

where $I_\mathrm{rad/conv}$ and $\mathbf{L}_\mathrm{rad/conv}$ are the moment of
inertia and the angular momentum vector of the core and envelope respectively,
$M_\mathrm{rad}$ and $R_\mathrm{rad}$ are the mass and radius of the radiative
core.

In addition to the change in internal structure, the angular velocity of the
convective surface of stars is also effected by the torques exerted by stellar
winds. The external torque leads to spin-down of the star at later stages of
evolution \citep{Schatzman,Irwin_Bouvier_2010,
Gallet_Bouvier_2013,Gallet_Bouvier_15}. To account for this spin-down POET
adopts a generalized formalism for the rate of angular momentum loss by assuming
winds apply torque to the convective zone in direction opposite to
$\mathbf{L}_{conv}$ with a magnitude:
\begin{equation}
    \label{eq:wind_torque}
    T_\mathrm{conv,wind} \equiv K\omega_\mathrm{conv}
    \mathrm{\min}(\omega_\mathrm{conv},\omega_\mathrm{sat})^2
    \left(\dfrac{R_{\star}}{R_\odot}\right)^{1/2}
    \left(\dfrac{M_{\star}}{M_\odot}\right)^{-1/2}
\end{equation}
where $K$ is the parameterized wind strength, $\omega_{conv}$ is the angular
frequency of the convective envelope, $\omega_{sat}$ is the frequency above
which the magnetic braking saturates, $M_\star$ and $R_\star$ are the mass and
radius of the star, and $M_\odot$ and $R_\odot$ are the present day mass and
radius of the Sun.

The orbital evolution model in POET is formulated in terms of the equilibrium
tide model, but as already pointed out in the introduction, dynamical tide
models can be incorporated by finding an effective phase lag.  The prescription
is based on the formalism of \citet{Lai_12}, but expanded to account for
eccentricity. In summary,

\begin{enumerate}
    \item The quadrupole moment of the tidal potential in a spherical
        $\left(\zeta, \theta, \phi\right)$-coordinate system is:
        \begin{equation}\label{eq:tidal_potential}
            U(\textbf{r},t)= - \sum_{m,m^{\prime}} U_{m,m^{\prime}} \zeta^2
            Y_{2,m}
            (\theta,\phi)\dfrac{a^3}{r^3(t)}\exp[-im^{\prime}\Delta\phi(t)]
        \end{equation}
        where a is the semi-major axis and r$\left(t\right)$ is the distance
        between the centers of the two objects at an arbitrary point on the
        orbit. $U_{mm^{\prime}}$ are the coefficients of the spherical
        harmonics.
    \item A Fourier series expansion is used to include effects of eccentric
        orbits:
        \begin{equation}
            \label{eq:fourier_expansion}
            \dfrac{a^3}{r^3(t)}\exp[-im^{\prime}\Delta\phi(t)] = \sum_s
            p_{m^{\prime},s}\exp(-is\Omega t)
        \end{equation}
        with the $p_{m^{\prime},s}$ coefficients pre-calculated as a function of
        eccentricity up to $s=400$ to allow for even extreme orbital
        eccentricities.  At e=0.9, truncating the tidal potential expansion to
        $s=200$ is accurate to better than 1 part in $10^5$. We doubled the
        maximum $s$ value for the sake of caution.

    \item Following the equilibrium theory of tides, the effects of tidal
        dissipation are introduced as a lag in the response of the fluid
        perturbation inside the star. Each $\left(ms\right)$-component of the
        tidal potential has a distinct tidal frequency ($\tilde{\omega}_{m,s} =
        s\Omega-m\Omega_s$) and thus a distinct tidal lag.  The Eulerian density
        perturbation and Lagrangian displacement is given as:
        \begin{align}
            \boldsymbol{\xi}_{m,s}(\mathbf{r},t)
            & =
            \dfrac{U_{m,s}}{\omega_0^2}
            \bar{\boldsymbol{\xi}}_{m,s}(\mathbf{r})
            \exp(-is\Omega t + i\Delta_{m,s})
            \label{eq:fluid}
            \\
            \delta\rho_{m,s}(\mathbf{r},t) & =
            \dfrac{U_{m,s}}{\omega_0^2}
            \delta\bar{\rho}_{m,s}(\mathbf{r})
            \exp(-is\Omega t + i\Delta_{m,s})
            \label{eq:density}
        \end{align}

        where $\delta\bar{\rho}_{m,s} =
        -\nabla(\rho\bar{\boldsymbol{\xi}}_{m,s})$ and
        $\omega_0\equiv\sqrt{GM/R^3}$ is the dynamical frequency of the primary
        star in the binary system.

    \item The individual $\left(ms\right)$-component of the tidal torque and
        rate of dissipation of energy are calculated as:
        \begin{align}
            \mathbf{T}_{m,s}
            &=
            \int d^3x
            \delta\rho_{m,s}(\mathbf{r},t)\mathbf{r}
            \times
            \left[-\nabla U^{\ast}(\mathbf{r},t)\right]
            \label{eq:torque}
            \\
            \dot{E}_{m,s}
            &=
            \int d^3x
            \rho_{m,s}(\mathbf{r})
            \dfrac{
                \partial\boldsymbol{\xi}_{m,s}(\mathbf{r},t)
            }{ \partial t
            }
            \left[-\nabla U^{\ast}(\mathbf{r},t)\right]
            \label{eq:E}
        \end{align}
    \item The evolution of the orbital parameters (semi-major axis $a$,
        eccentricity $e$, angle between spin angular momentum and orbital
    angular momentum vectors $\theta$, and angular frequency of primary star
$\omega$) in terms of torque, energy, and angular momentum is calculated using
the following equations:

        \begin{align}
            \dot{a} &= a\dfrac{-\dot{E}}{E} \\
            \dot{e} &=
            \dfrac{2(\dot{E}L+2E\dot{L})L(M+M^{\prime})}{G(MM^{\prime})^3} \\
            \dot{\theta} &= \dfrac{(T_z - \tilde{T}_z)\sin\theta}{L} -
            \dfrac{(T_x - \tilde{T}_x)\cos\theta}{L} -  \dfrac{T_x +
            \mathcal{T}_x}{S} \\
            \dot{\omega} &=  \dfrac{(T_y +
            \mathcal{T}_y)\cos\theta}{L\sin\theta} + \dfrac{T_y +
            \mathcal{T}_y}{S\sin\theta}
        \end{align}
        where  $M$ is the mass of tidally distorted star, $M^{\prime}$ is the
        mass of the companion raising the tides, $T_x, T_y, T_z$ are the
        components of tidal torque along $x, y, z$-directions, $\mathcal{T}_x,
        \mathcal{T}_y, \mathcal{T}_z$ are the non-tidal torques (i.e. stellar
        wind and core-envelope coupling), and $\tilde{T}_x, \tilde{T}_y,
        \tilde{T}_z$ are the tidal torques on the orbit due to other zones.  The
        contribution of both stars to the orbital evolution is included.
\end{enumerate}

The relation between the tidal quality factor and the phase lag is:

\begin{equation}
    \label{eq:tidal_lag_relation}
    \Q = \dfrac{15}{16\pi\Delta_{\star}^{\prime}}
\end{equation}

The tidal lag introduced in the above method allows for an arbitrary
parameterization of tidal dissipation depending on the model being studied.  For
equilibrium tide models with constant phase lag, $\Q$ has a fixed value, while
for for dynamical tide models or variable phase lag equilibrium models, $\Q$
will in general depend on the orbital and stellar properties, as well as the
particular tidal potential term being dissipated. In this work, we restrict our
investigation to stars with similar internal structure, and only allow for a
dependence on tidal frequency.

%% file: data.tex
\section{Input Data}
\label{sec:input_data}

\subsection*{Stellar Rotational Periods}
\label{sec:rotation_periods}

We wish to infer the infer the parameters of the $Q_{\ast}^{\prime}$ model using
the observed rotation period of the primary star in binary star systems. We
model the spin of the star as a Gaussian distribution whose mean and standard
deviation are estimated as described below.

\citet{Lurie_2017} performed an extensive study of the light curves of the
eclipsing binaries available in the \textit{Kepler Eclipsing Binary Catalog} and
estimated the rotation period of the primary stars showing starspot modulations
of their brightness.

The authors analyzed data available from \textit{Kepler} Data release 25 in
quarters 0-17. The Presearch Data Conditioning (PDC) pipeline suppresses stellar
variability with $P \gtrsim  20\,\text{days}$. Hence, rotation periods are
available only for stars with $P < 20 \text{ days}$.

The morphology of stellar variability for single stars is a well-modeled
phenomenon. The authors use the out-of-eclipse region of the light curves to
analyze the variability of stars in eclipsing binaries. Depending on the
variations, they classify the light curves into six categories by visual
inspection: starspot modulations, ellipsoidal modulations, pulsations, other
periodic variability (do not belong to previous three or unidentified Heartbeat
stars), non-periodic out-of-eclipse (flat-light curves or long-term smooth
variation due to instrumental artifacts) and not eclipsing binary (lack of clear
eclipses). Out of these, only starspot modulations can provide information about
the rotation period of the stars. Starspot modulations are quasi-sinusoidal,
with differential rotation and the formation and dissipation of starspots
introducing long-term modulation \citep{Davenport2015}. The authors identify 816
eclipsing binaries that show such variations.

Two common ways to detect periodicity of variability in a time-variable signal
are auto-correlation function (ACF) and Lomb-Scargle Periodogram
\citep{lombscargle1976}. Following \citet{McQuillan_2013}, \citet{Lurie_2017}
identifies an initial rotation period for each binary using ACF. Since ACF is
not sensitive to multiple rotation period peaks in the light curve signal, it
can only serve as a validation for signals with multiple periods present due to
the differential rotation of the star. For this purpose, the authors turn to
Lomb-Scargle Periodograms in order to detect more than one peak of periodic
variability in the out-of-eclipse light curves using a procedure by
\citet{Rappaport2014}, with ACF used for validation.

From the Lomb-Scargle Periodogram, the authors identified the two most
significant peaks, imposing a threshold that peaks must be at least $30\%$ of
the highest peak. Each peak is then classified into two groups. Lowering this
threshold results in the detection of spurious signals which might not be
related to stellar variability. A window of frequency is selected for each group
by selecting the local minima in a smooth periodogram on the right and left of
the dominant peak. Inside this window, they again use the same threshold to
identify the next dominant subpeak. Hence, for each system, there are a maximum
of four possible solutions for the rotation period. Table 2
of~\citet{Lurie_2017} shows the periods and their respective peak height along
with the period and peak height detected by the ACF. The ACF period comes
closest to the most dominant peak in the periodogram.

As the authors themselves mention, the rotation period corresponding to the
highest peak obtained in the first group ($P_{1\min}$ in Table 2
of~\cite{Lurie_2017}) will be the closest to the equatorial rotation period. We
adopt this as the nominal value of the rotation period of the primary star
in our analysis. To obtain the uncertainty on the rotation period, we select the
rotation period corresponding to the subpeak in the same group ($P_{1\max}$ in
Table 2 of~\cite{Lurie_2017}) and calculate the difference between $P_{1\max}$
and $P_{1\min}$. See Sec. \ref{sec:reliable_constraints} for an additional
consideration we apply to account for the possibility that this procedure may
under-estimate the rotational period uncertainty.

\subsection*{Binary Parameters}

\begin{table}
    \centering
    \caption{The maximum likelihood parameters of the binary systems used in
    this analysis and their spins.}
    \label{tab:system_parameters}
    \resizebox{0.49\textwidth}{!}{%
        \input{systems_maxlike_table}
    }
\end{table}

\citet{Windemuth_2019} (W19 from now on) inferred the values of orbital and
stellar parameters for detached eclipsing binaries in the \textit{Kepler} field
from the available photometric data. The authors selected 2877 eclipsing and
ellipsoidal binaries available in the Villanova \textit{Kepler} EB Catalogue
\citep{Prsa2011, Kirk2016}. In order to get a catalog of well constrained
detached eclipsing binaries, they first selected binaries with eclipse depths
for the primary and secondary stars to be $>5\%$ and $>0.1\%$ of the normalized
flux to ensure high signal to noise ratio. Next, they used the morphology
parameter, provided by VKEBC, as $\texttt{morph}<0.6$ to ensure the binaries in
the sample are detached. The morphology parameter is based on a scheme that
classifies light curves based on the type of binary star observed. Low values of
this parameter correspond to a well-detached binary exhibiting clear separation
between the eclipses, while higher values correspond to over-contact binary
systems with sinusoidal variations \citep{Matijevi2012}.

The orbital parameters and the stellar parameters of these systems are inferred
by performing a simultaneous fit for the Light Curve (LC) and the Spectral
Energy Distribution (SED). The observational uncertainties are then propagated
to the model parameters using MCMC simulations with a likelihood function based
on how well the LC, SED, extinction and distance are simultaneously reproduced.
The authors performed this fit for 22 parameters. For our analyses, only the
following parameters are of interest:
\begin{itemize}
    \item Sum of masses of the primary and secondary star $\left(M_{sum}\right)$
    \item Ratio of secondary mass to the primary mass  $\left(q\right)$
    \item Age of the system $\left(\tau\right)$
    \item Metallicity $\left(\left[\text{Fe}/\text{H}\right]\right)$, assumed to
        be the same for both stars
    \item Eccentricity components of the orbit $\left(e\cos{\omega},
        e\sin{\omega}\right)$.
\end{itemize}

W19 provided posterior samples for 728 binaries. We restrict this sample
further. First, since our analysis relies on the spin period of the primary
star, we perform a cross-match between these 728 binaries and binaries available
in \citet{Lurie_2017}. We also apply limits on the stellar masses and
metallicity ($ 0.4 M_{\odot} < M_{\star} < 1.2 M_{\odot} $ and $-1.014 <
\left[\text{Fe}/\text{H}\right] <0.537$), which further reduced our dataset.
The mass limits were imposed to select only stars that have similar main
sequence internal structure to the Sun, namely a significant convective outer
shell, but not fully convective. The metallicity limits are driven by
limitations imposed by POET, which at the moment is only capable of
interpolating the stellar evolution within these mass and metallicity limits.
Finally, we require reasonable orbital eccentricity $e<0.8$ and orbital period
of no more than 20 days.

Our final dataset contains 70 binary systems. We use the publicly available
thinned posterior samples for our parameters of interest as observed probability
distributions for Bayesian analysis. Table \ref{tab:system_parameters} gives the
maximum likelihood values (as reported by W19) of the mass of the primary star
$\left(M_1\right)$, mass of the secondary star $\left(M_2\right)$, and
eccentricity components of the orbit ($e\sin\omega$, $e\cos\omega$), along with
the orbital period of the system, rotation period of the primary star and the
$1\sigma-$uncertainty assumed for the spin period (see Sec.
\ref{sec:rotation_periods}). Note that for each binary only a single spin period
is measured, which we assume is that of the primary (brighter) star.

%% file: systems_maxlike_table.tex
\begin{tabular}{cccccccc}
    \hline
    \textbf{KIC}
    & $\mathbf{M}_{\mathbf{1}} [M_\odot]$
    & $\mathbf{M}_{\mathbf{2}} [M_\odot]$
    & $\mathbf{e}\textbf{sin}\boldsymbol{\omega}$
    & $\mathbf{e}\textbf{cos}\boldsymbol{\omega}$
    & $\mathbf{P_{orb}} [\mathrm{days}]$
    & $\mathbf{P_\star} [\mathrm{days}]$
    & $\boldsymbol{\sigma}_{\mathbf{P_\star}} [\mathrm{days}]$\\
    \hline

    6962018
    & 0.75
    & 0.69
    & 0.0002
    & 0.00024
    & 1.27
    & 1.3
    & 0.003\\%

    11616200
    & 1.16
    & 0.67
    & -0.014
    & 0.00019
    & 1.72
    & 1.7
    & 0.003\\%

    7798259
    & 0.73
    & 0.60
    & -0.039
    & 7.6e-07
    & 1.73
    & 1.7
    & 0.019\\%

    4380283
    & 1.01
    & 0.88
    & -4.9e-06
    & 7.8e-05
    & 1.74
    & 1.7
    & 0.003\\%

    7732791
    & 0.64
    & 0.65
    & 0.039
    & 0.00025
    & 2.06
    & 2.0
    & 0.021\\%

    4579321
    & 0.81
    & 0.61
    & -0.03
    & -8.6e-05
    & 2.11
    & 2.1
    & 0.03\\%

    11200773
    & 1.09
    & 0.42
    & -0.2
    & 0.00037
    & 2.49
    & 2.5
    & 0.033\\%

    9656543
    & 0.79
    & 0.76
    & -0.0041
    & 8.4e-05
    & 2.54
    & 2.5
    & 0.13\\%

    3834364
    & 1.03
    & 0.51
    & -0.15
    & 0.00025
    & 2.91
    & 2.9
    & 0.025\\%

    11228612
    & 0.99
    & 0.70
    & 0.0044
    & 4.2e-05
    & 2.98
    & 2.9
    & 0.007\\%

    6312521
    & 1.02
    & 0.76
    & -0.091
    & 0.00016
    & 3.02
    & 3.0
    & 0.045\\%

    10960995
    & 0.94
    & 0.83
    & -5.9e-08
    & 1.2e-07
    & 3.12
    & 3.0
    & 0.11\\%

    11147276
    & 1.18
    & 0.89
    & -0.021
    & -6e-06
    & 3.13
    & 3.1
    & 0.054\\%

    10091257
    & 1.00
    & 1.06
    & 0.021
    & 1.4e-05
    & 3.4
    & 3.3
    & 0.06\\%

    6525196
    & 0.91
    & 0.85
    & -5.4e-05
    & 4.7e-05
    & 3.42
    & 3.3
    & 0.085\\%

    5022440
    & 1.10
    & 0.73
    & 4.7e-07
    & -1.1e-06
    & 3.69
    & 3.6
    & 0.13\\%

    5802470
    & 1.13
    & 0.77
    & 2.7e-06
    & 2.2e-05
    & 3.79
    & 3.8
    & 0.073\\%

    4815612
    & 1.17
    & 1.00
    & -5.3e-08
    & -1.6e-07
    & 3.86
    & 3.7
    & 0.083\\%

    3241344
    & 1.01
    & 0.51
    & -0.0055
    & -0.0041
    & 3.91
    & 3.9
    & 0.014\\%

    11403216
    & 1.07
    & 0.45
    & -0.036
    & 0.00034
    & 4.05
    & 4.0
    & 0.019\\%

    10935310
    & 0.84
    & 0.54
    & 0.17
    & -0.0018
    & 4.13
    & 3.9
    & 0.014\\%

    10031409
    & 1.16
    & 1.11
    & 0.004
    & 6e-06
    & 4.14
    & 4.0
    & 0.17\\%

    7838639
    & 1.21
    & 1.08
    & -0.0084
    & 9.2e-06
    & 4.23
    & 4.1
    & 0.081\\%

    3973504
    & 0.88
    & 0.57
    & -0.1
    & -4.9e-05
    & 4.32
    & 4.3
    & 0.032\\%

    8957954
    & 1.03
    & 1.01
    & -2.5e-08
    & -1.9e-08
    & 4.36
    & 4.3
    & 0.086\\%

    11252617
    & 0.88
    & 0.47
    & 0.24
    & 3.6e-05
    & 4.48
    & 4.5
    & 0.05\\%

    4285087
    & 0.99
    & 0.97
    & -2e-05
    & -5.5e-05
    & 4.49
    & 4.5
    & 0.049\\%

    4346875
    & 0.94
    & 0.55
    & -0.022
    & -9.8e-06
    & 4.69
    & 4.8
    & 0.11\\%

    4947726
    & 1.14
    & 0.53
    & -0.15
    & 0.079
    & 4.73
    & 4.4
    & 0.21\\%

    11233911
    & 0.84
    & 0.94
    & 0.016
    & 0.00018
    & 4.96
    & 4.8
    & 0.076\\%

    12004679
    & 0.92
    & 0.88
    & -0.0093
    & 0.0018
    & 5.04
    & 5.6
    & 0.21\\%

    2860788
    & 1.03
    & 0.83
    & -0.0045
    & -6.1e-06
    & 5.26
    & 5.0
    & 0.084\\%

    9971475
    & 1.10
    & 0.80
    & -0.00033
    & 4.6e-05
    & 5.36
    & 2.0
    & 0.012\\%

    5730394
    & 1.17
    & 0.95
    & -0.019
    & -0.00025
    & 5.52
    & 6.3
    & 0.38\\%

    5181455
    & 0.96
    & 0.54
    & -1.7e-06
    & -3.6e-07
    & 5.58
    & 5.5
    & 0.11\\%

    8381592
    & 1.01
    & 0.73
    & -0.064
    & -0.0044
    & 5.78
    & 5.7
    & 0.088\\%

    8618226
    & 1.08
    & 0.70
    & 1.5e-05
    & 0.0002
    & 5.88
    & 6.0
    & 0.25\\%

    3838496
    & 0.99
    & 1.07
    & -0.0056
    & 0.00079
    & 5.98
    & 5.7
    & 0.061\\%

    10385682
    & 1.00
    & 0.99
    & -0.015
    & -5.9e-06
    & 6.21
    & 6.1
    & 0.048\\%

    8580438
    & 0.98
    & 0.43
    & 0.00062
    & 0.002
    & 6.5
    & 7.5
    & 0.19\\%

    10965963
    & 1.08
    & 1.00
    & 0.045
    & -0.036
    & 6.64
    & 6.2
    & 0.41\\%

    8746310
    & 0.80
    & 0.47
    & -2.5e-06
    & 6.7e-05
    & 6.86
    & 8.3
    & 0.4\\%

    7362852
    & 1.16
    & 1.11
    & 0.0036
    & -0.0015
    & 7.04
    & 7.1
    & 0.24\\%

    8543278
    & 0.82
    & 0.71
    & 0.023
    & 0.013
    & 7.55
    & 8.6
    & 0.51\\%

    6927629
    & 1.08
    & 0.97
    & 1.4e-05
    & 2.7e-05
    & 7.74
    & 8.9
    & 0.27\\%

    8364119
    & 0.92
    & 0.88
    & 0.016
    & 0.018
    & 7.74
    & 9.0
    & 0.22\\%

    6949550
    & 0.88
    & 0.88
    & -0.011
    & -0.26
    & 7.84
    & 6.9
    & 0.39\\%

    3348093
    & 0.65
    & 0.68
    & 0.26
    & 0.0069
    & 7.96
    & 8.2
    & 0.27\\%

    9532123
    & 0.99
    & 0.75
    & -0.097
    & -0.21
    & 8.21
    & 7.3
    & 0.37\\%

    9892471
    & 0.91
    & 0.56
    & -0.013
    & -0.00029
    & 8.27
    & 9.4
    & 0.29\\%

    7816680
    & 1.18
    & 0.96
    & -0.018
    & 1.9e-05
    & 8.59
    & 11.1
    & 0.21\\%

    11391181
    & 0.87
    & 0.82
    & 0.077
    & 0.17
    & 8.62
    & 7.9
    & 0.28\\%

    4839180
    & 1.08
    & 0.86
    & -0.0021
    & -0.0025
    & 8.85
    & 11.0
    & 1.3\\%

    5652260
    & 0.88
    & 0.74
    & 0.001
    & 0.0012
    & 8.93
    & 11.0
    & 1.2\\%

    11232745
    & 0.73
    & 0.65
    & 3.6e-08
    & -3.9e-08
    & 9.63
    & 12.7
    & 0.75\\%

    8984706
    & 1.04
    & 1.03
    & -0.013
    & 0.015
    & 10.1
    & 12.5
    & 1.3\\%

    5393558
    & 0.92
    & 1.04
    & 0.23
    & -0.023
    & 10.2
    & 9.4
    & 0.48\\%

    9353182
    & 1.14
    & 0.57
    & -0.042
    & 0.036
    & 10.5
    & 10.5
    & 0.22\\%

    4352168
    & 1.03
    & 0.73
    & -0.15
    & -0.15
    & 10.6
    & 10.2
    & 0.6\\%

    7336754
    & 1.14
    & 0.73
    & 0.016
    & 0.012
    & 12.2
    & 13.2
    & 1\\%

    6029130
    & 0.92
    & 0.87
    & 0.0099
    & 0.011
    & 12.6
    & 17.3
    & 0.33\\%

    5871918
    & 0.67
    & 0.47
    & -0.32
    & -0.15
    & 12.6
    & 12.4
    & 0.16\\%

    11704044
    & 0.84
    & 0.82
    & 1e-06
    & 2.7e-06
    & 13.8
    & 19.4
    & 1.1\\%

    6359798
    & 0.86
    & 0.79
    & 0.079
    & -0.41
    & 14.2
    & 3.4
    & 0.029\\%

    10936427
    & 1.08
    & 1.12
    & -0.0074
    & 0.0016
    & 14.4
    & 16.6
    & 1.1\\%

    4678171
    & 0.68
    & 0.59
    & 0.0025
    & -0.0035
    & 15.3
    & 20.3
    & 5.4\\%

    7987749
    & 0.90
    & 0.79
    & 0.043
    & -0.14
    & 17
    & 17.5
    & 0.45\\%

    8356054
    & 0.77
    & 0.71
    & 0.37
    & 0.027
    & 17.1
    & 16.7
    & 2.3\\%

    10330495
    & 1.03
    & 0.52
    & -0.056
    & -0.04
    & 18.1
    & 21.1
    & 2.3\\%

    10992733
    & 0.94
    & 0.83
    & 0.17
    & 0.34
    & 18.5
    & 15.6
    & 0.27\\%

    10711913
    & 0.88
    & 0.76
    & 0.28
    & -0.077
    & 19.4
    & 18.6
    & 1.1\\%

    4252226
    & 0.99
    & 0.92
    & 0.13
    & -0.47
    & 21.9
    & 11.9
    & 0.76\\%

    10215422
    & 1.12
    & 0.72
    & 0.0014
    & 0.29
    & 24.8
    & 24.8
    & 2.2\\%

    4773155
    & 1.03
    & 0.95
    & -0.36
    & 0.26
    & 25.7
    & 19.0
    & 2\\%
	\hline
\end{tabular}

%% file: q_formalism.tex
\section{$\Q$ Formalism}
\label{sec:q_formalism}

In this work, we assume that the tidal dissipation occurs only in the convective
zones of the stars, meaning that we assume the tidal torque is applied only to
the convective zone, with the core spin affected only indirectly through the
core-envelope coupling.  As mentioned in Section~\ref{sec:introduction}, the
effects of the binary system parameters on tidal dissipation can be studied by
using a suitable parameterization for the tidal lag, $\Delta_{m,m^{\prime}}$ or
the corresponding tidal quality factor, $Q_{m,m^{\prime}}^{\prime}$, such that
it can capture the orbital dynamics of the system under the effects of tides.

As discussed in Sec. \ref{sec:tidal_model}, we allow each Fourier component of
the tidal potential to experience its own effective lag. In this work, we
restrict the possibilities to some extent, by assuming that
$\Delta_{m,m^{\prime}}^{\prime}$ depends solely on the tidal frequency (Eq.
\ref{eq:tidal_frequency}). Each equilibrium or dynamical tidal model predicts a
different frequency dependence. In this work, we further simplify the
prescription to assume that $\Delta_{m,m^{\prime}}$ has a power-law dependence,
on $\Omega_{\text{tide}} = m\Omega_{\text{orbit}} - m'\Omega_{\star}$.
Furthermore, in order to avoid clearly non-physical and numerically unstable
behavior occurring if $\Delta_{m,m'} \rightarrow \infty$, we assume the
dissipation saturates to some maximum value:
\begin{equation}
    \label{eq:lag_formalism}
    \Delta_{m,m^{\prime}}^{\prime}
    =
    \mathrm{sign}(\Omega_{m,m^{\prime}}) \Delta_0
    \min\left[
        1,
        \left| \dfrac{\Omega_{m,m^{\prime}}}{\Omega_{\text{break}}}
        \right|^{\alpha}
    \right]
\end{equation}
or in terms of $\Q$:
\begin{equation}
    \label{eq:q_formalism}
    Q_{m,m^{\prime}}^{\prime}
    =
    \mathrm{sign}(P_{m,m^{\prime}}) Q_0
    \max\left[
        1,
        \left| \dfrac{P_{m,m^{\prime}}}{P_{\text{break}}} \right|^{\alpha}
    \right]
\end{equation}
where $\alpha$ is the powerlaw index parametrizing the frequency dependence,
$\Delta_0=\frac{15}{16\pi Q_0}$ parametrizes the maximum allowed dissipation,
$\Omega_{m,m'}=\frac{2\pi}{P_{m,m'}}$ is the tidal frequency of a particular
tidal wave, and $\Omega_{\text{break}}=\frac{2\pi}{P_\text{break}}$ is the
frequency at which the dissipation saturates.  See Appendix \ref{sec:q_tweak}
for a small modification we introduce, which improves numerical performance
without significantly changing the evolution.

We now wish to find constraints on $Q_0$, $\alpha$, and $P_{\text{break}}$ that
can simultaneously reproduce the observed spins of the primary stars in all the
binaries in our input sample. In practice, the tidal effects on the spin of each
binary will be dominated by a narrow range of frequencies, so for a given
binary, we will be able to constrain $Q_{m,m^{\prime}}^{\prime}$ for a narrow
range of periods. Furthermore, because tides get weaker as the separation
between the two stars increases, we expect that for the lowest periods of our
sample of binaries all systems will be synchronized, thus providing only a lower
limit on the dissipation (upper limit on $\Q$), while at the longest periods
stellar spins will not be significantly affected by tides, providing only an
upper limit to the dissipation (lower limit on $\Q$). Because the effects of
tides decrease very precipitously as the orbit gets wider, there should be only
a small number of binaries in between these two regimes providing two-sided
constraints. It should be noted that the range of orbital periods with
asynchronous binaries whose spin is nonetheless significantly affected by tides
is much wider than one may naively expect. This is caused by the fact that the rate
at which stars lose angular momentum due to magnetized winds scales as the cube
of the angular velocity. Hence, maintaining synchronous rotation requires ever
stronger tidal coupling as the orbital period decreases. Even with that, however,
combining the constraints from multiple binary systems is required to get a
measurement of the tidal dissipation and detect any potential
frequency-dependence.

%% file: bayesian_analysis.tex
\section{Bayesian Analysis}
\label{sec:bayesian_analysis}

We use the Markov Chain Monte Carlo algorithm as implemented in the Python
package \texttt{emcee} \citep{Foreman_Mackey_2013}. \texttt{emcee} is based on a
class of ensemble sampler algorithms that uses affine invariant transformation
as a proposal function to simultaneously advance multiple not-independent chains,
referred to as walkers, in the parameter space. We refer the reader to
\citep{Foreman_Mackey_2013} and references therein for details of the algorithm.

The choice of the number of walkers is a compromise between parallelization
efficiency and the need to accumulate a sufficient number of steps to ensure the
distribution of samples has converged to the target posterior. For our analysis,
we use 64 walkers for each binary to sample the 9-dimensional space of
parameters required to calculate the orbital evolution.

The parameters sampled by MCMC and their respective priors are listed in Table.
\ref{tab:sampling_parameters}. For the physical parameters of the binaries we
construct a joint prior using Kernel density estimation from the MCMC samples
published by W19, and for the dissipation parameters we use very broad uniform
priors.

% \usepackage{graphicx}
\begin{table}
    \centering
    \caption{The parameters sampled during MCMC and their priors. The W19 joint
    prior is constructed using KDE from the W19 MCMC samples.}
    \label{tab:sampling_parameters}
    \resizebox{0.49\textwidth}{!}{%
        \begin{tabular}{cccc}
            \hline
            \textbf{Parameter} & \textbf{Units} & \textbf{Description} &
            \textbf{Prior} \\
            \hline
            $M_1$ & $M_{\odot}$ & Mass of the primary star & W19 \\
            $M_2$ & $M_{\odot}$ & Mass of the secondary star & W19 \\
            $\left[\frac{\mathrm{Fe}}{\mathrm{H}}\right]$ & - & Metallicity of
            both stars & W19 \\
            $\tau$ & Gyr & Age of the system & W19 \\
            $e$ & - & Present day eccentricity of the Orbit & W19 \\
            $P_{\star,\text{init}}$ & days & Initial spin period of the stars &
            $U()$ \\
            $\log_{10}Q_0$ & - & Dissipation parameter. See Eq.
            \ref{eq:q_formalism} & $U(5, 10)$ \\
            $\alpha$ & - & Dissipation parameter. See Eq.  \ref{eq:q_formalism}
            & $U(-5,5)$ \\
            $\log P_\text{break}$ & rad/day & Dissipation parameter. See Eq.
            \ref{eq:q_formalism} & $U\left(\log 0.5\,d, \log 50\,d \right)$ \\
            \hline
        \end{tabular}%
    }
\end{table}

\subsection{Prior Transformation}
\label{sec:prior_transform}

Because the prior distributions built from W19 samples are quite complex
structures in a multi-dimensional space, the MCMC sampling becomes very
inefficient. In order to remedy this situation, instead of directly sampling
from the prior distribution of the parameters from Table
\ref{tab:sampling_parameters}, we use a transformed set of parameters ($u_i$
with $i=1\ldots 9$) such that each of those has an independent prior
distribution uniformly distributed between 0 and 1. Given a sample of values for
$u_i$, the physical and dissipation parameters required to calculate the
evolution are then found by applying a prior transformation. This procedure
dramatically simplifies the posterior likelihood the MCMC process must sample
from, making it directly proportional to the likelihood that the actual spin of
the primary star as observed today is equal to the spin period predicted by the
evolution assuming the given parameters.

The joint prior distribution we wish to impose on the physical parameters of
each binary (the first 5 parameters listed in Table \ref{tab:sampling_parameters})
is the posterior distribution W19 samples are drawn from. We start by isolating
only the parameters we need. For each sample, we combine the components of
eccentricities ($e\cos{\omega}$ and $e\sin{\omega}$) to just eccentricity ($e =
\sqrt{(e\cos\omega)^2 + (e\sin\omega)^2}$). We then construct a marginalized
probability distribution function $\pi \left(M_{sum}, q, \tau,
[\mathrm{Fe}/\mathrm{H}], e\right)$, where $M_{sum} \equiv M_1 + M_2$ and $q
\equiv M_1/M_2$, using a Kernel density estimator (KDE) with a Gaussian kernel
with a bandwidth determined using the improved Sheather-Jones algorithm
\citep{sheather2010}.

Finally, W19 samples were generated assuming independent $U(0,1)$ priors on
$e\cos\omega$ and $e\sin\omega$. This has the undesirable effect that when
converted to prior on eccentricity, the probability of $e=0$ is zero, even if
the cloud of points in $e\cos{\omega}$, $e\sin{\omega}$ space clearly includes
the origin. This will propagate to the prior distribution we impose resulting in
erroneously imposing an upper limit on the dissipation (lower limit on $\Q$) for
such systems. To see this, consider a system for which the primary star is
consistent with spinning synchronously with the orbit, and the $e\cos{\omega}$,
$e\sin{\omega}$ cloud of points is centered on the origin (i.e.  observations
are consistent with circular orbit). We expect that for such a system
arbitrarily small $\Q$ should be acceptable since large amounts of tidal
dissipation will predict a completely circularized and synchronized system.
Instead, the prior on eccentricity imposed by W19 will exclude that
configuration, erroneously requiring a small eccentricity comparable to the
uncertainty in $e\sin{\omega}$ (the much less well constrained of the two
components) to survive to the present day. Instead of the W19 prior, we wish to
impose independent priors $e \in U(0,1)$ and $\omega \in U(0,2\pi)$. The desired prior
is obtained from the W19 prior by simply dividing by $e$ and re-normalizing.

To be precise, if the $s^{\text{th}}$ W19 sample has values for the parameters given by
$M_{sum}=M_s$, $q=q_s$, $\tau=\tau_s$, $\feh = \feh_s$, and
$e=e_s$, the prior probability density of the systems parameters is given by:
\begin{align}
    \pi & \left(M_{sum}, q, \tau, \feh, e\right) \propto \nonumber \\
    &~~~~~~~~\sum_s k_M(M_{sum}-M_s) k_q(q - q_s)~~~\times \nonumber \\
    &~~~~~~~~~~~~~~k_\tau(\tau-\tau_s)  k_\feh\left(\feh - \feh_s\right) \frac{k_e(e-e_s)}{e}
\end{align}
%
%\begin{equation}
%
%    \begin{array}{c}
%
%        \pi\left(M_{sum}, q, \tau, \feh, e\right) \\
%
%        \propto\\
%
%        \sum_s k_M(M_{sum}-M_s) k_q(q - q_s) k_\tau(\tau-\tau_s)
%        k_\feh\left(\feh - \feh_s\right) \frac{k_e(e-e_s)}{e}
%
%    \end{array}
%
%\end{equation}
%
where $k_M\ldots$ are the kernels used for each of the quantities, and the
factor of $1/e$ in the last term is responsible for changing the priors as
explained above.

The resulting prior transformation converting $u_1, \ldots, u_5$ for a given
MCMC sample to the corresponding physical system parameters is then:

\begin{equation}
    \begin{array}{rcl}
        M_{sum}  & = & F_M^{-1} (u_1)\\
        q_i  & = & F_q^{-1} (u_2|M_{sum})\\
        \tau_i  & = & F_\tau^{-1} (u_3|M_{sum},q)\\
        Z_i  & = & F_\feh^{-1} (u_4|M_{sum},q,\tau)\\
        e_i  & = & F_e^{-1} (u_5|M_{sum},q,\tau,\feh)\\
    \end{array}
\end{equation}
with
\begin{equation}
    \begin{array}{r@{\ }c@{\ }l}
        F_M(M) & \equiv & \int_{-\infty}^{M} \text{d}M_{sum}
        \int_{-\infty}^{\infty} \text{d}q \int_{-\infty}^{\infty} \text{d}\tau
        \int_{-\infty}^{\infty} \text{d}\feh \int_{-\infty}^{\infty} \text{d}e\\
        && \quad \quad \pi\left(M_{sum}, q, \tau, \feh, e\right)\\
        F_q(q|M_{sum}) & \equiv & \int_{-\infty}^{q} \text{d}q'
        \int_{-\infty}^{\infty} \text{d}\tau \int_{-\infty}^{\infty}
        \text{d}\feh \int_{-\infty}^{\infty} \text{d}e\\
        &&\quad \quad \pi\left(M_{sum}, q', \tau, [\mathrm{Fe}/\mathrm{H}],
        e\right)\\
        && \ldots
%%
%%
%%
%        F_\tau(\tau|q,M_{sum}) & \equiv & \int_{-\infty}^{\tau} \text{d}\tau'
%        \int_{-\infty}^{\infty} \text{d}[\mathrm{Fe}/\mathrm{H}]
%        \int_{-\infty}^{\infty} \text{d}e \\
%%
%        && \quad \quad \pi\left(M_{sum}, q, \tau', [\mathrm{Fe}/\mathrm{H}],
%        e\right)\\
%%
%%
%%
%        F_\feh(\feh|\tau,q,M_{sum}) & \equiv & \int_{-\infty}^{\feh}
%        \text{d}\feh' \int_{-\infty}^{\infty} \text{d}e \pi\left(M_{sum}, q,
%        \tau, \feh', e\right)\\
%
%
    \end{array}
\end{equation}
The above calculations are feasible because calculating multi-dimensional
integrals is not necessary. Kernel functions are normalized, so integrals over
$\pm\infty$ are all unity.

The priors on the dissipation parameters and the initial spin frequency of the
primary star are assumed uniform. The corresponding prior transform for those is
just a simple shift and scaling to match the limits.

The parameters not listed in Table \ref{tab:sampling_parameters} and required by the
evolution (Sec. \ref{sec:tidal_model}) are assumed constant with the following
values:
\begin{equation}
    \begin{array}{lcl}
        K & =& 0.17\,
        M_\odot\,R_\odot^2\,\mathrm{day}^2\,\mathrm{rad}^{-2}\,\mathrm{Gyr}^{-1}\\
        \omega_{sat} & = & 2.45\,\mathrm{rad}\,\mathrm{day}^{-1}\\
        \tau_{c-e} & = & 5\,\mathrm{Myr}.
    \end{array}
\end{equation}

\subsection{MCMC Convergence Diagnostic}
\label{sec:convergence}

The samples generated by an MCMC algorithm can be shown to follow the specified
posterior distribution in the limit of infinitely long chains. Real-world
applications then need to demonstrate that the generated chains are long enough
to get a good approximation to the distribution. There are two concerns that
must be addressed. First, in infinite chains the starting positions are
irrelevant. In finite chains however, an MCMC process requires some number of
steps before subsequent samples can be shown to come from the desired
distribution to a good approximation. This is usually referred to as the burn-in
period, and a typical practice is to discard the early samples. The second
concern is that one needs a sufficient number of post-burn-in samples in order to
estimate the targeted quantities with a desired precision.

In this work, we are interested in estimating quantiles of the posterior
distribution. \citet{raftery1991many} derived diagnostics for single chain MCMC
of a single quantity that answer the questions:
\begin{itemize}
    \item What is the smallest $N$ such that the probability that the $N+1$ step
        in the chain is below some threshold value is within a specified
        precision of the limit of that probability for $N\to\infty$.
    \item What is the variance of the estimated value of the cumulative
        distribution at the threshold value from the MCMC chain after the $N^{\text{th}}$
        sample.
\end{itemize}

The \citet{raftery1991many} procedure cannot be directly applied to
\texttt{emcee} chains because that involves multiple non-independent chains. In
\citet{Patel2022} and \citet{Penev2022} we adapted the \citet{raftery1991many} procedure
for \texttt{emcee} chains, allowing us to estimate when enough samples have been
generated to reliably and precisely estimate a specified quantile of some target
quantity.

In this work, we select a grid of tidal periods $P_{tide,i}$ and use the
\texttt{emcee} samples of $Q_0$, $\alpha$, and $P_{\text{break}}$ to evaluate
the tidal model (Eq. \ref{eq:q_formalism}) at each of these periods to obtain
samples of $Q'_i = Q_{m,m^{\prime}}^{\prime}(P_{tide,i})$. For each of these
quantities, we wish to find the 2.3\%, 15.9\%, 84.1\%, and 97.7\% quantiles.
Using the adapted \citet{raftery1991many} formalism, we find a burn-in period
for each $Q'_i$  for each quantile, requiring that the fraction of the first
post-burn-in samples of $Q'_i$ (one for each walker) below the quantile is
within $10^{-3}$ of the equilibrium probability, and estimate the uncertainty in
the CDF for each $Q'_i$ and quantile combination from the remaining samples
after discarding the burn-in.

Because finding a given quantile requires knowing the burn-in period, and
finding the burn-in period requires knowing the quantile, we iterate between the
two to find a mutually consistent combination.

%% file: constraints.tex
\section{$Q_{\star}^{\prime}$ Constraints}
\label{sec:constraints}

\subsection{Individual Constraints}

From the MCMC we obtain posterior samples for $Q_0$, $P_{\text{break}}$ and
$\alpha$ for each system. These samples are converted to $\Q(P_\text{tide})$
using Equation~\ref{eq:q_formalism} evaluated at 30 separate tidal periods
$\log-$uniformly spaced between 1 and 50 days. For each tidal period we apply
the convergence diagnostics described in Section~\ref{sec:convergence} to find
the 2.3, 15.9, 84.1, and 97.7 percentiles of $\Q$ at that tidal period
(corresponding to $\pm1\sigma$ and $\pm2\sigma$ uncertainties) together with the
burn-in steps and cumulative distribution uncertainty.

\begin{figure*}
    \begin{center}
        \includegraphics[width=\textwidth]{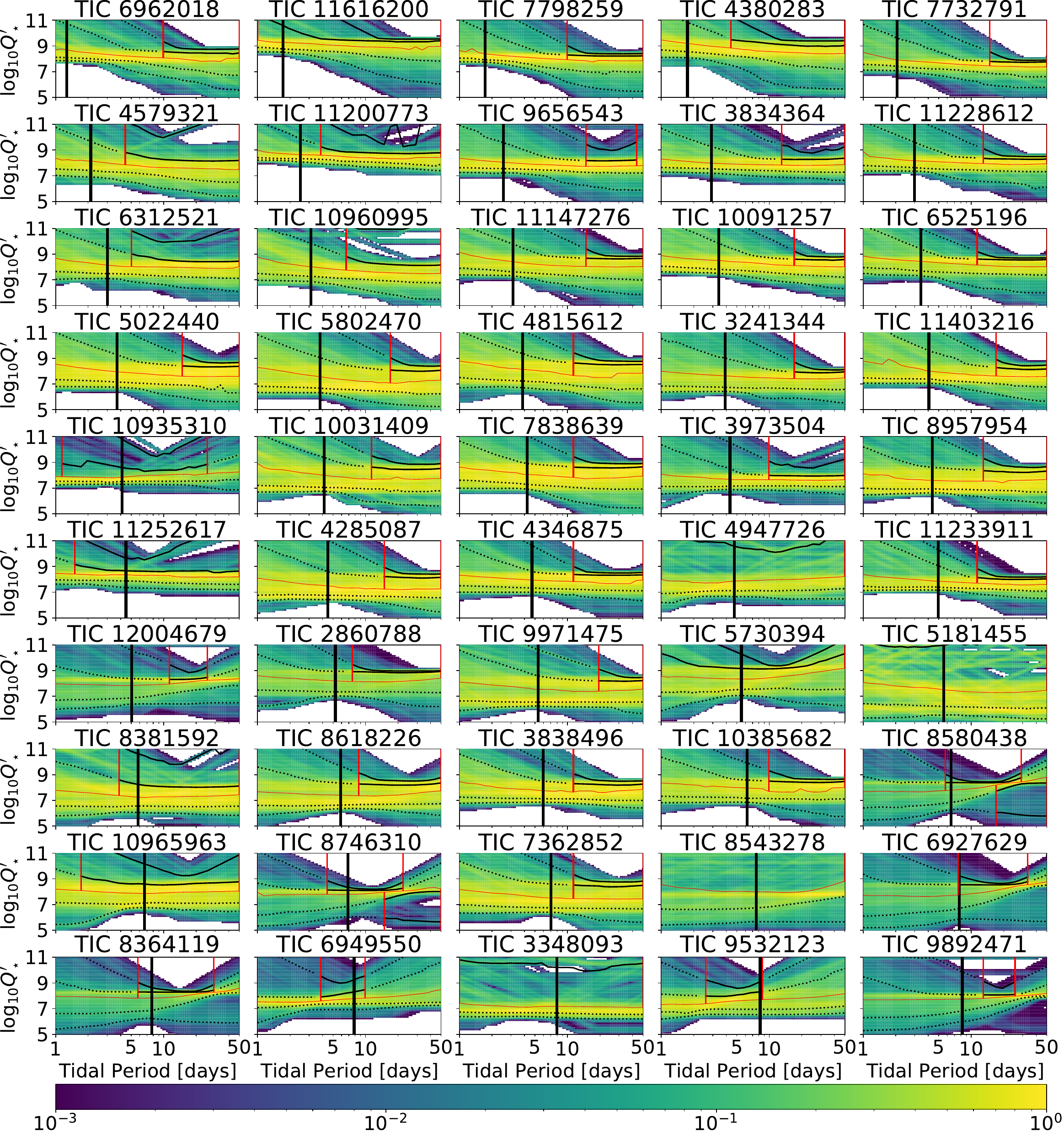}
    \end{center}
    \caption{
        Individual constraints on $\log_{10}\Q$ as function of tidal periods,
        $P_{tide} \in[0.5,50]$ days for each binary systems. The color heat map
        represents the posterior probability density function of $\Q$ at each
        tidal period generated by evaluating Eq. \ref{eq:q_formalism} for each
        MCMC sample after the estimated burn-in period. The horizontal black
        lines show the 2.3, 15.9, 84.1, and 97.7 percentiles with solid line
        where we believe the constraints are reliable and dotted where we
        believe they are either significantly affected by our assumed priors, or
        may be affected by under-estimated uncertainty of the spin (see Sec.
        \ref{sec:reliable_constraints}). The thick black vertical line shows the
        orbital period of the system, the thin red horizontal line is the median
        and vertical red lines delineate the range of periods outside of which
        the assumed priors could be significantly driving the constraints.
    }
    \label{fig:ind_const_1}
\end{figure*}

\begin{figure*}
    \begin{center}
        \includegraphics[width=\textwidth]{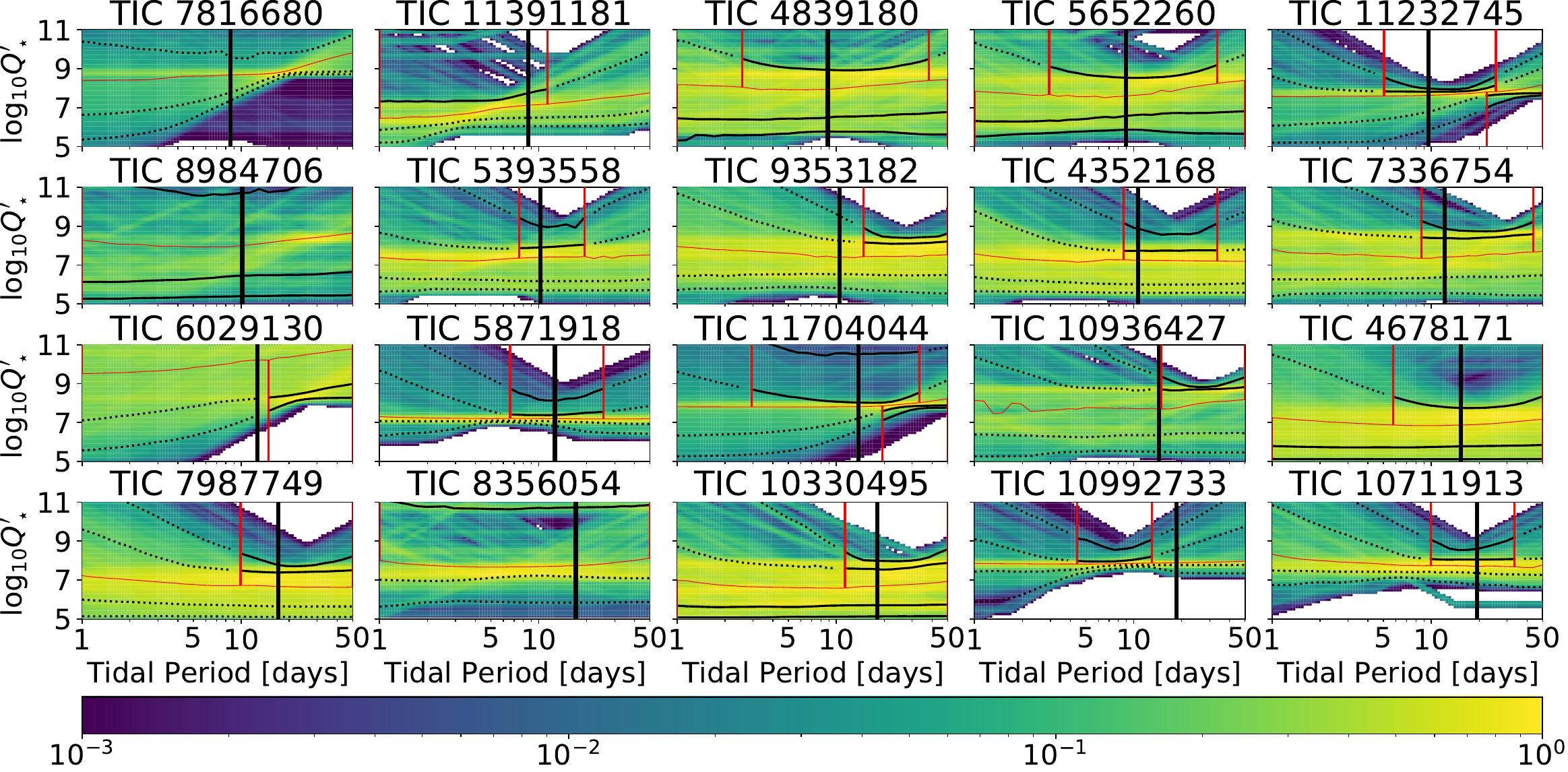}
    \end{center}
    \caption{
        Same as Fig. \ref{fig:ind_const_1} but for the remaining systems.
    }
    \label{fig:ind_const_2}
\end{figure*}

Figures \ref{fig:ind_const_1} and \ref{fig:ind_const_2} show the KDE estimate
of the posterior distribution of $\Q$ at each tidal period (color-coded heat
map), the 2.3, 15.9, 84.1, and 97.7 percentiles (black dotted or solid curves),
the $50^{\text{th}}$ percentile as the thin horizontal red curve, and the orbital period
(thick vertical black line). The solid parts of the quantile curves, delineated
with vertical red lines, mark the range of periods for which the $\Q$
distribution and quantiles are reliable, i.e. dominated by the data rather than
the priors and not affected by potentially under-estimated uncertainty of the
spin (see Sec. \ref{sec:reliable_constraints} for details).

\begin{figure*}
    \begin{center}
        \includegraphics[width=\textwidth]{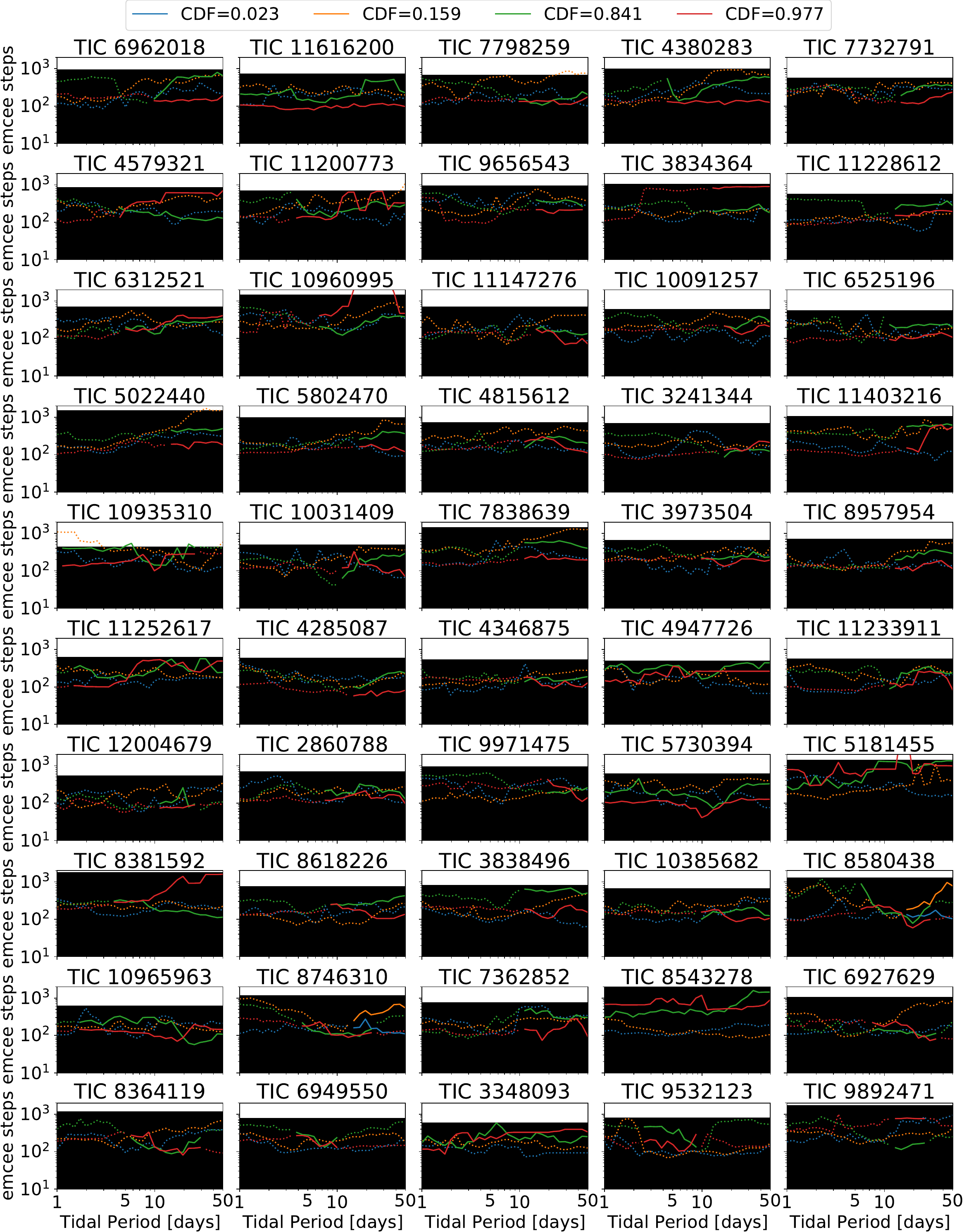}
    \end{center}
    \caption{
        Estimated burn-in periods required for each of the percentiles of
        $\log_{10}\Q(P_{tide})$ at each tidal period (lines). Solid lines are
        used for the range of periods where the constraints are deemed reliable
        (Sec. \ref{sec:reliable_constraints}), dotted outside this range. The
        black area shows the number of steps accumulated for each system.
    }
    \label{fig:burnin_1}
\end{figure*}

\begin{figure*}
    \begin{center}
        \includegraphics[width=\textwidth]{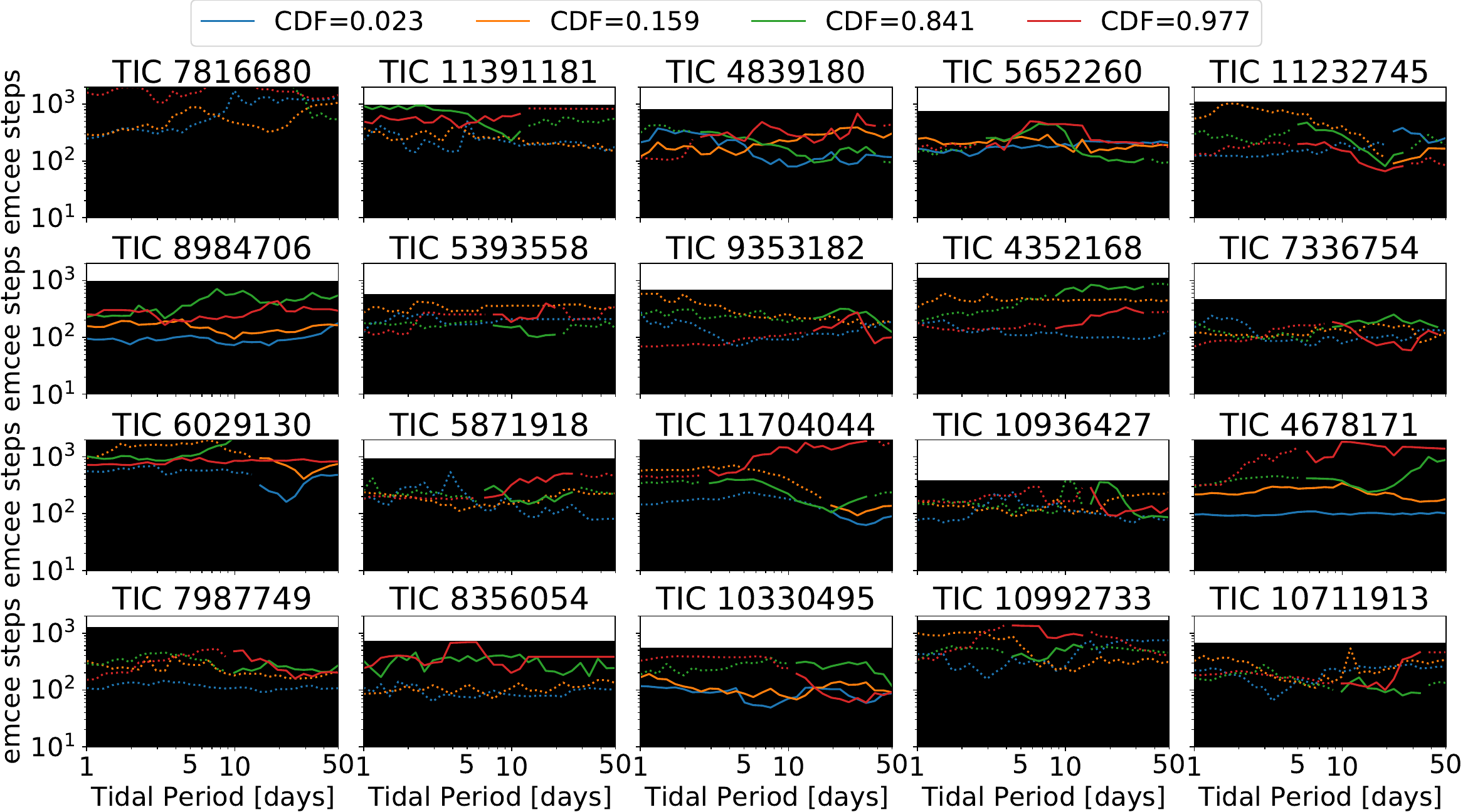}
    \end{center}
    \caption{
        Same as Fig. \ref{fig:burnin_1} but for the remaining systems.
    }
    \label{fig:burnin_2}
\end{figure*}

Figures \ref{fig:burnin_1} and \ref{fig:burnin_2} show the estimated
burn-in periods (see Sec. \ref{sec:convergence}) for the four quantiles of $\Q$
at each tidal period as the curves (solid corresponding to the reliable range
of periods, dotted otherwise), as well as the total number of steps we
accumulated for each system (black area). As that figure demonstrates, for the
tidal periods where a given quantile is deemed reliable (see Sec.
\ref{sec:reliable_constraints}), for all our systems we have accumulated enough
samples to move past the burn-in period.

\begin{figure*}
    \begin{center}
        \includegraphics[width=\textwidth]{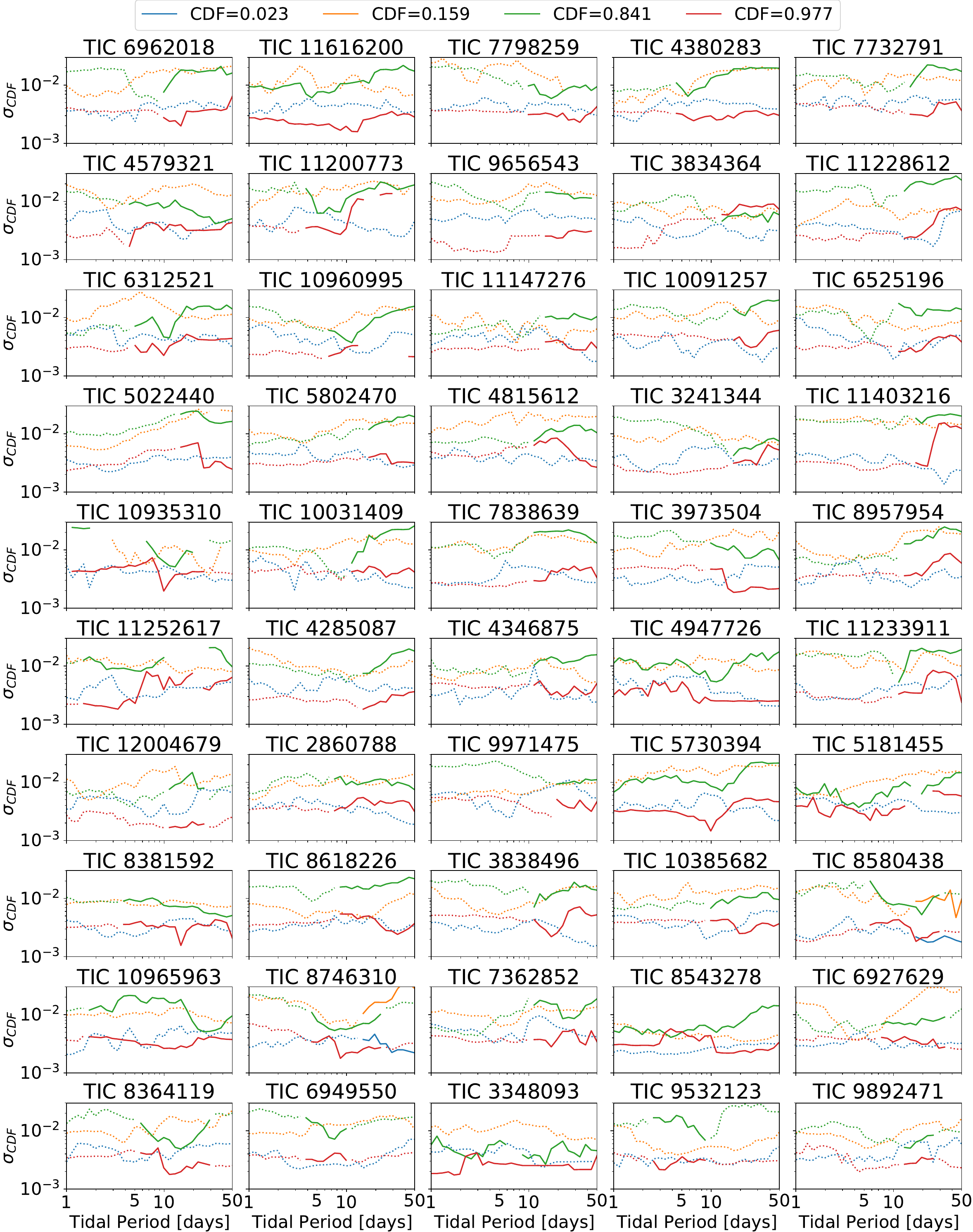}
    \end{center}
    \caption{
        Estimated uncertainty (standard deviation) of the fraction of the
        distribution of $\log_{10}\Q(P_{tide})$ that lies below each quantile.
        Solid lines are used for the range of periods where the constrained are
        considered reliable (see Sec. \ref{sec:reliable_constraints}), dotted
        lines otherwise.
    }
    \label{fig:cdfstd_1}
\end{figure*}

\begin{figure*}
    \begin{center}
        \includegraphics[width=\textwidth]{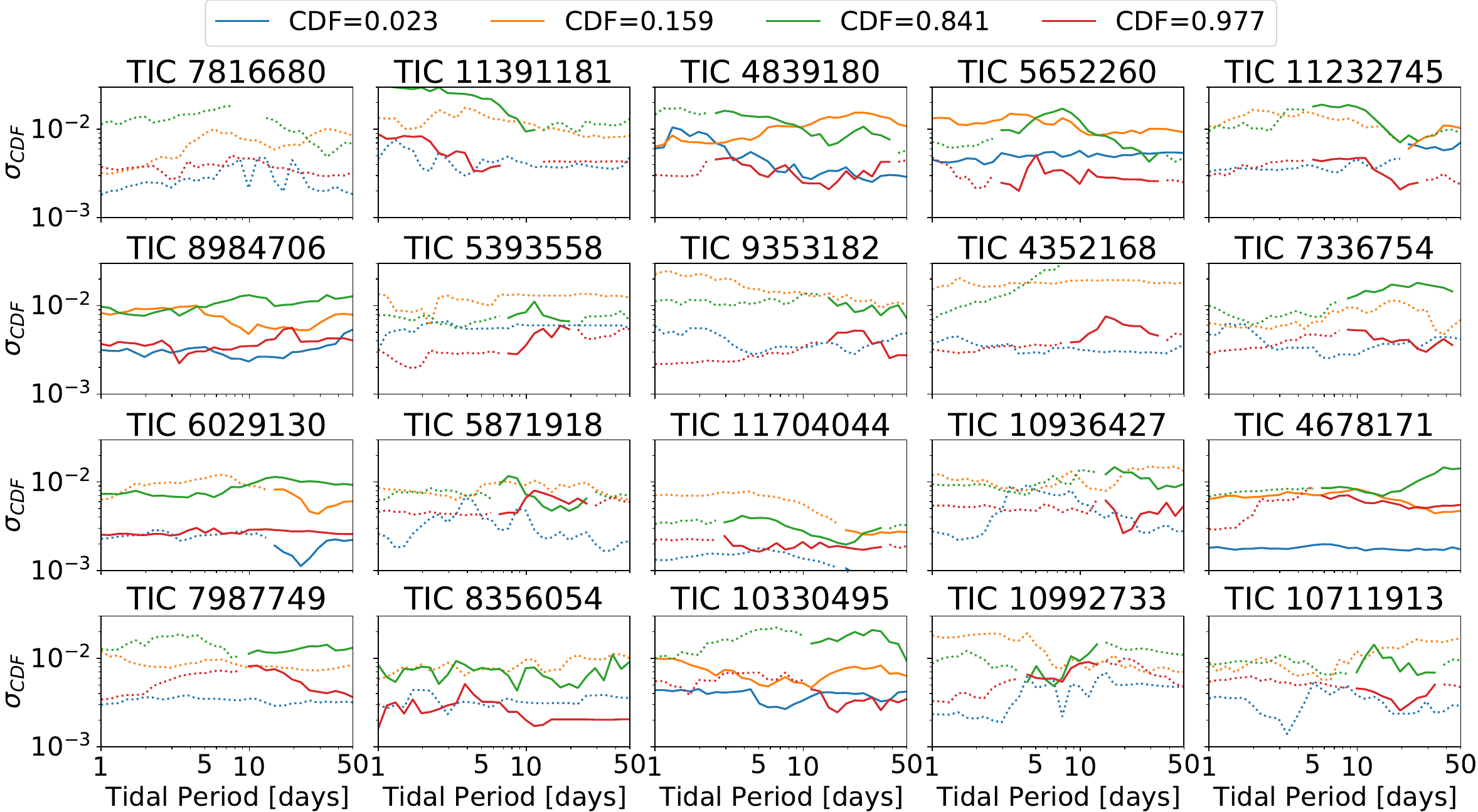}
    \end{center}
    \caption{
        Same as Fig. \ref{fig:cdfstd_1} but for the remaining systems.
    }
    \label{fig:cdfstd_2}
\end{figure*}

Finally, Figures \ref{fig:cdfstd_1} and \ref{fig:cdfstd_2} show the estimated
standard deviation of the cumulative distribution function (CDF) at each
quantile estimated as described in Sec. \ref{sec:convergence} from the samples
after burn-in is discarded.

\subsection*{Two-sided limits on $\Q$}

The systems for which we obtain two-sided limits significantly differ between
the orbital period and the spin period of the primary star and significantly
deviate from the spin period of an isolated star given the same mass and age.
For example, in the case of \texttt{KIC11232745}, the orbital period of the
system is 9.6 days, while the spin period is 12.7 days. The lower bound on
$Q_{\star}^{\prime}$ comes from not allowing high dissipation, which would cause
the primary star to synchronize with the orbit. The upper bound on
$Q_{\star}^{\prime}$ is due to the requirement of minimal dissipation such that
the tidal influence on the spin of the star is not negligible compared to the
effect of stellar winds. Given the very steep drop off of the effects of tides
with orbital separation (tidal torque decreases as the fourth power of the
orbital period), at first glance it may appear that such systems should be so
rare that finding several in a sample of 70 binaries would be surprising.
However, two-sided limits on $\Q$ correspond to stars with spin periods
comparable (though not quite equal) to the orbital period. As a result, since
stellar wind torques scale strongly with the spin period (third power of the
spin period), which is itself related to the orbital period, the transition from
fully synchronized systems to systems unaffected by tides is much more gradual
than thinking about tides alone would predict.

\subsection*{Upper Limit on $\Q$}

There are systems where we only obtain an upper bound on $\log_{10}\Q$ (lower
bound on the dissipation). In these cases, the spin period of the primary star
is synchronized with the orbit. Above this limit, the dissipation is not high
enough to maintain a spin-orbit lock and keep the star synchronized with the
orbit. However, reproducing the observed state of the system can tolerate
arbitrarily large dissipation (arbitrarily small $\Q$) since that will just
result in the spin-orbit lock being achieved earlier.

\subsection*{Lower Limit on $\Q$}

Similarly, there are systems with only a lower bound on $\log_{10}\Q$ (an upper
bound on the dissipation). This limit is obtained for systems whose spins are
consistent with that of isolated stars of similar mass and age. The inability to
distinguish the spin from that of an isolated star is generally driven by the
fact that the age is usually the least well constrained parameter.  We stress
that the non-tidal parameters we assume (core-envelope coupling timescale, wind
strength, wind saturation frequency etc.) were tuned to reproduce the observed
spins of isolated stars with well known ages (ones residing in open clusters).
Hence, if we calculate the evolution of a binary under the assumption of
$Q_\star' \rightarrow \infty$, the predicted spin of each star in the binary is
the same as that of an isolated star with the same mass and age. The lower limit
to $Q_\star'$ produced by these systems is thus a statement that if the
dissipation were any larger than this, it would have had a detectable effect on
the spin.

For some systems, the lower bound on $\log_{10}{Q_{\star}^{\prime}}$ can be due
to eccentricity rather than the spin period of the primary star. It may be that
the uncertainty in the age does not allow us to distinguish the spin from that
of an isolated star, but if non-zero orbital eccentricity is detected with high
significance, dissipation must be weak enough to avoid circularizing the system.

\subsection{Reliable Portion of Individual $Q_{\star}^{\prime}$ Constraints}
\label{sec:reliable_constraints}

Examining the posterior distributions of $\log_{10}\Q$ in Fig.
\ref{fig:ind_const_1} and \ref{fig:ind_const_2}, it is clear that each
individual binary is sensitive to the dissipation only in a relatively narrow
range of periods. For example, the difference between the median and the 97.7$^{\text{th}}$
percentile of the distribution has a sharp minimum. Sufficiently far away from
that period, the increase of the 97.7$^{\text{th}}$ percentile is limited to large
extent by the fact that we restrict the powerlaw index $\alpha$ in Eq.
\ref{eq:q_formalism} to a maximum value of 5. Similarly, for some systems, the
small quantiles peak at some period, and, away from that, the slope is limited by
the assumption that $\alpha>5$.

In order to select the region where constraints are dominated by the observed
state of the binary rather than the priors, we evaluate the absolute difference
between the highest/lowest quantiles and the median at each tidal period. We
then select the period range where this difference remains within 0.5 dex of its
minimum value. The upper/lower half of the distribution is then only deemed
reliable within that period range.

For the systems for which we obtain a lower bound or two-sided bound on
$\log_{10}\Q$, we must further examine the difference in the measured spin
period and the orbital period. Generally, the statistical uncertainty in
measuring the spin period is small compared to stellar differential rotation. As
described in Sec.  \ref{sec:rotation_periods}, we estimate that difference using
the detection of multiple peaks in the Lomb-Scargle periodogram. However, if the
star spots modulating the lightcurve of a particular star happen to all come
from a narrow rang of latitudes at the time they were observed, our approach
will under-estimate the amount of differential rotation. The result could be a
star which synchronized the spin to the orbit, but for which, due to
differential rotation, the LC modulation happens to have a slightly different
period and the difference appears significant because we under-estimate the
amount of differential rotation. In turn, this will produce an upper limit to
the dissipation (lower limit on $\Q$) that is not justified by the observations.
To compensate for this possibility, we trust only the upper limit to
$\log_{10}\Q$ our analysis gives for systems with measured spin periods within
17\% of the measured orbital period (approximately the amount of differential
rotation between the poles and equator of the Sun) and ignore the lower limit.
Suggestively, the distribution of fractional difference between spin and orbital
period for our systems has a gap between 17\% and 21\%, with most systems below
17\%, perhaps hinting at a switch from a population of synchronized to
non-synchronized systems, though the sample is not large enough to claim this is
statistically significant or to explore possible explanations.

%We selected posterior samples for the orbital and stellar parameters provided
%by \citet{Windemuth_2019} as distribution functions for bayesian analysis.
%\citet{Windemuth_2019} used uniform priors, $U(0,1)$ on the two eccentricity
%components: $e\cos{\omega}, e\sin{\omega}$ in their analysis. We transformed
%these samples to eccentricity using the relation: $e = \sqrt{{e\cos{\omega}}^2
%+ {e\sin{\omega}}}^2$ and sampled from the probability distribution as shown in
%Equations. This methodology requires a correction in the probability
%distribution of eccentricity by multiplying it by a factor $\dfrac{1}{e}$.
%Otherwise, there is a higher preference for non-zero eccentricities in the
%probability distribution. Avoiding this correction can result in a lower limit
%on $Q_{\star}^{\prime}$ for circular systems, which comes from non-zero
%eccentricity rather than synchronization. To correct for this, we visually
%inspect the posterior samples of $e\cos{\omega} \text{ vs } e\sin{\omega}$ for
%each system and identify the system where zero eccentricity is within the
%spread of the two distributions (Figure). Next, from this subset, we identify
%the systems that show high likelihood at low $Q_{\star}^{\prime}$ values. We
%discard the lower limit for these systems and renormalize the probability
%distribution for $Q_{\star}^{\prime}$ at each tidal period.

\subsection{Combined Constraints}
As expected, individual constraints are sensitive to a limited range of
frequencies, and most binaries only produce upper or lower limits, but only a
few produce both. As a result, meaningful measurement of $\Q$ can only be
obtained by combining the individual constraints. Furthermore, since we selected
all our binaries to have similar internal structure, it is not unreasonable to
assume they should have similar dissipation.

A common constraint, which agrees with all the constraints obtained by the
individual binary systems, can be found by multiplying the posterior probability
density functions of $\Q(P_\text{tide})$ for the binary systems for each
$P_\text{tide}$, obtained using KDE. This of course is complicated by the
considerations described in Sec. \ref{sec:reliable_constraints}. For a given
system, at a given tidal period, we have already identified whether we trust
both halves of the distribution, just the upper half, just the lower half, or
neither. If neither side of the distribution is reliable, we simply do not
include that system in the product for the given tidal period. If only one side
of the distribution is reliable, we replace the unreliable part of the
probability density with the probability density at the median and re-normalize
before including the distribution in the product for the combined distribution.

One of our binaries -- KIC 7816680 -- produces constraints that are in conflict
with the combined constraint from the other binaries for all tidal periods above
7 days. An outlier binary can have many explanations:
\begin{enumerate}
    \item There may be additional objects in the system resulting in a very
        different orbital evolution than we calculate assuming just two objects
    \item A third object in the system or another star along the same line of
        sight could be contributing extra light to the system. Since W19 models
        the spectral energy distribution assuming just the two binary components
        that could shift the inferred properties of the binary.
    \item Spin period measurements sometimes detect an alias of the spin period,
        or perhaps the spot modulations come from the secondary star instead of
        the primary, invalidating our assumption that measured spin is that of
        the brighter star.
    \item The stellar evolution models used in the W19 analysis to find binary
        parameter distributions and in our orbital evolution calculations may
        not be applicable. In fact, W19 already report discrepancies for young
        stars and caution discrepancies are also expected for post-main-sequence
        stars, and POET stellar evolution interpolation is unreliable past the
        end of the main sequence as it was never designed to handle such stars.
\end{enumerate}

We suspect KIC 7816680 falls in the final category. The primary star has a
maximum likelihood mass of $1.2M_\odot$ and maximum likelihood age of 9.4\,Gyr,
hence the W19 model of this binary involves a post-MS primary, so some amount of
difficulty with stellar evolution models is expected.

\begin{figure}
    \centering
    \includegraphics[width=\columnwidth]{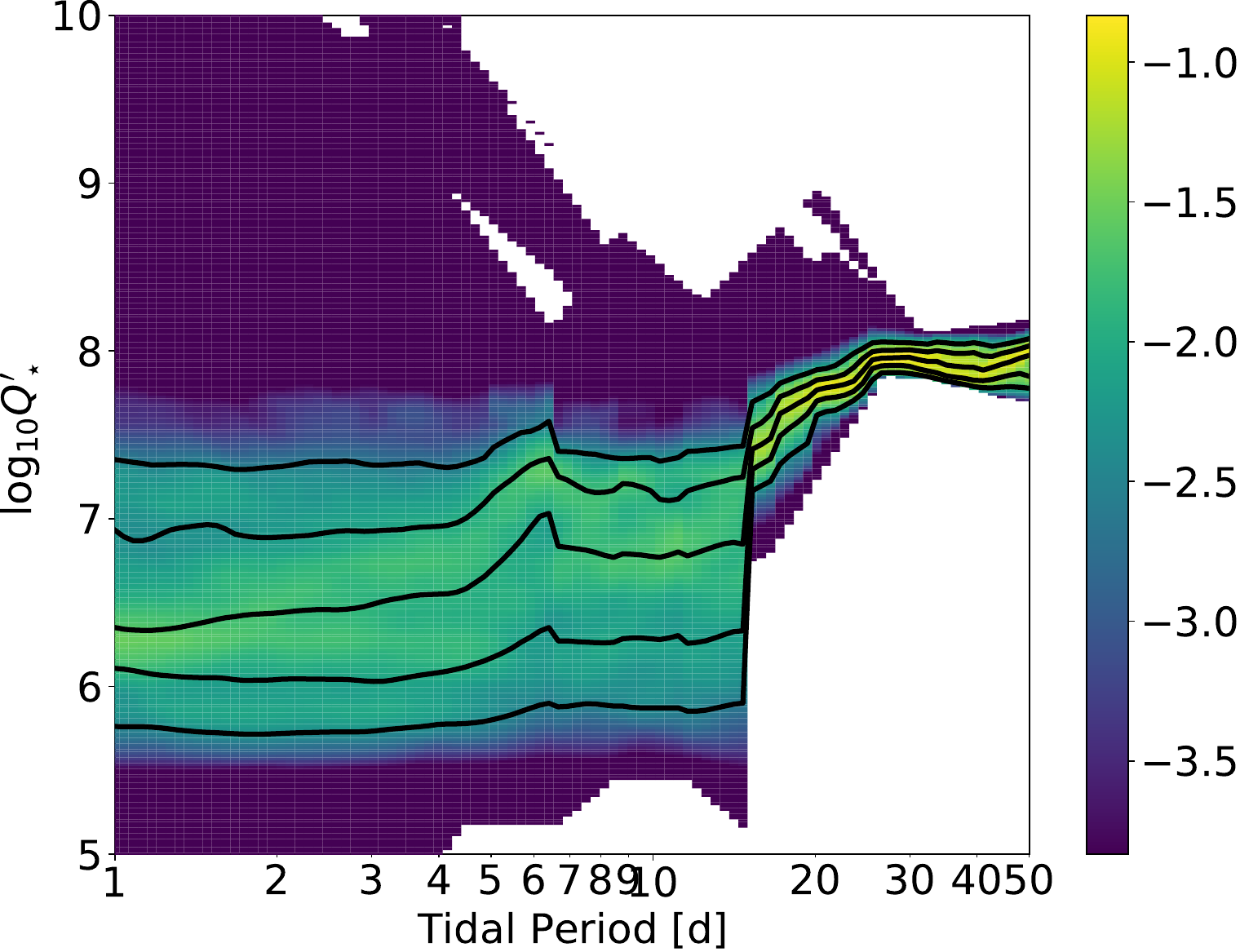}
    \caption{
        Combined constraints on $Q_{\star}^{\prime}$ as a function of tidal
        period, calculated by multiplying the probability distribution functions
        obtained for the individual binary system at each tidal period, after
        applying corrections per the discussion in Sec.
        \ref{sec:reliable_constraints}. The black lines correspond to the 2.3,
        15.9, 50.0, 84.1, and 97.7 percentiles of $\log_{10}\Q$ constraints.
    }
    \label{fig:combined}
\end{figure}

Fig. \ref{fig:combined} shows the combined constraint on $\log_{10}\Q$ obtained
through the above procedure, excluding KIC 7816680.

\begin{figure}
    \centering
    \includegraphics[width=\columnwidth]{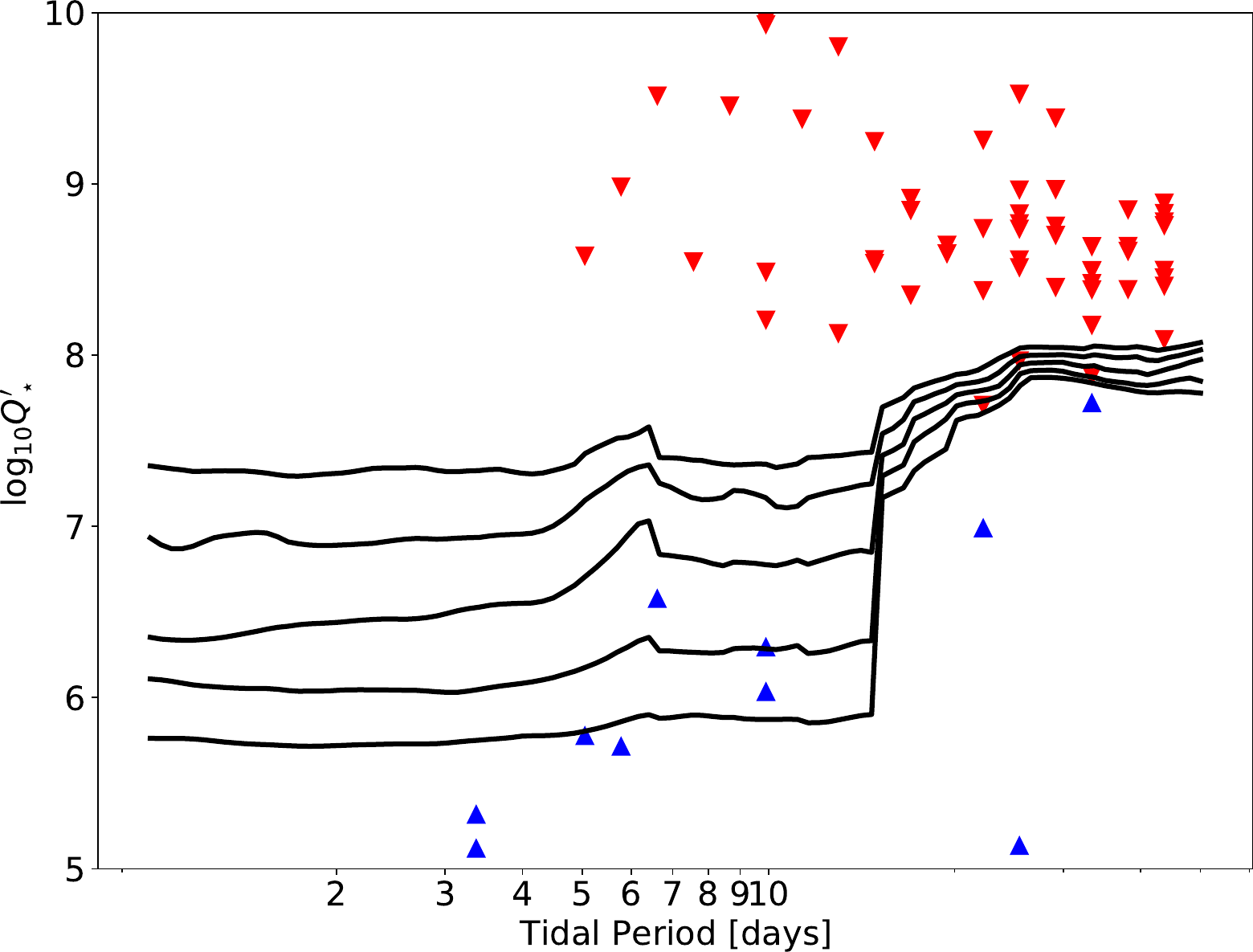}
    \caption{
        Comparison between the combined constraint and the individual
        constraints. The lines show the 2.3, 15.9, 50.0, 84.1, and 97.7
        percentiles of the combined $\log_{10}\Q$ constraints (same as in Fig.
        \ref{fig:combined}). The blue up and red down triangles show the
        2.3 and 97.7 percentiles of the constraint for individual systems at the
        tidal period where the difference between the two is smallest.
    }
    \label{fig:combined_vs_individual}
\end{figure}

An important question that must be considered is whether the combined
constraints are indeed consistent with all individual constraints. After all, it
is possible to multiply two mutually highly inconsistent distributions and
obtain a combined constraint that is somewhere in the middle that is highly
inconsistent with either distribution. Fig.~\ref{fig:combined_vs_individual}
shows the 2.3, 15.9, 50.0, 84.1, and 97.7 percentiles of the combined
constraints (same as the lines in Fig. \ref{fig:combined}) along with the 2.3
and 97.7 percentiles for each individual system at the tidal period where the
distance between the 2.3 and 97.7 percentiles is smallest. As can be seen in
Fig.~\ref{fig:combined_vs_individual}, there is at least some overlap between
the allowed $\log_{10}\Q$ values for each system and the combined constraint. In
other words, the combined constraint is capable of reproducing the observed
spins of all the binaries in our sample except KIC 7816680.

%% file: discussion.tex
\section{Discussion}
\label{sec:discussion}

\subsection{Merging with Prior Results}
\label{sec:comparison}

\begin{figure}[h!]
    \begin{center}
        \includegraphics[width=0.45\textwidth]{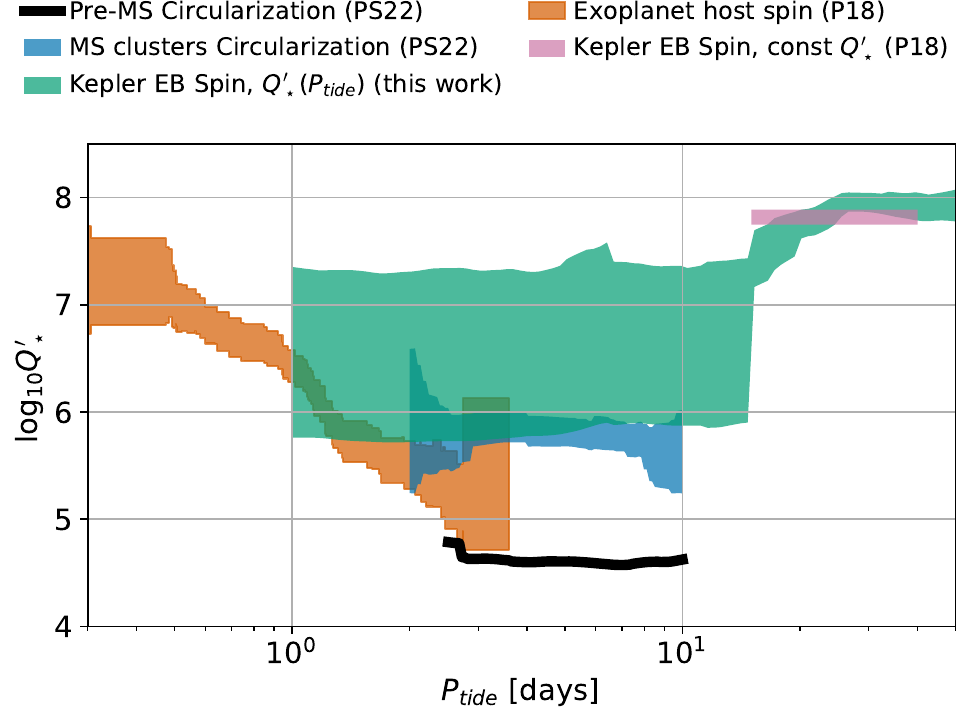}
    \end{center}
    \caption{
        Constraints on tidal dissipation ($\log_{10}\Q$) obained by our group
        using similar analysis to this article. See text for brief descriptions.
    }
    \label{fig:merged_constraints}
\end{figure}

Our group has embarked on a systematic effort to construct a comprehensive
empirical picture of tidal dissipation in low mass stars by combining
constraints derived from multiple tidal effects observable in multiple
collections of binary star or exoplanet systems. We rely on a common approach of
analyzing each system individually, employing Bayesian analysis to account for
observational and model uncertainties, and combining the individual constraints
to find a common prescription for $\Q$ capable of explaining the observed
properties of each population of objects. We begin by placing this result in the
context of the previous results from our group (Fig.
\ref{fig:merged_constraints}).

The most closely related study by our group \citep{Patel2022} (PP22 from now on)
analyzed the tidal effects on the spin of 41 of the Kepler eclipsing binaries
analyzed here, but assuming $\Q = $ constant and without the benefit of the W19
parameters, which were not available at the time. We used a similar approach,
analyzing one binary at a time, and then combining the individual posterior
distributions of $\Q$ for each binary system. We obtained a common constraint
$\Q = 7.82\pm0.035$, which was in agreement with each of the individual
constraints. As can be seen from Fig.~\ref{fig:merged_constraints}, there is
also agreement with the results found by this study. That is of course expected,
since many of the same binaries are analyzed both here and by \citep{Patel2022},
though with less precise parameters and using a different prescription for $\Q$.

\citet{Penev2022} (PS22 hereafter) studied tidal circularization of the single
line spectroscopic binaries in three open clusters (M35, NGC 6819, and NGC 188)
to obtain constraints of a frequency dependent $\Q$. PS22 constructed an
envelope in the period-eccentricity distribution for each cluster which
encompasses all observed binary systems, and required that dissipation is strong
enough to place any binary below the envelope and weak enough to allow
reproducing the observed present day eccentricity of each system. Using the same
formalism for frequency-dependent $\Q$ as defined in
Equations~\ref{eq:q_formalism}, PS22 found that for the two older clusters
($\sim$2\,Gyr old NGC 6819, and $\sim$7\,Gyr old NGC 188) $5.5 < \log_{10}\Q <
6$ for $2\,\text{days} < P_{\text{tide}} < 10\,\text{days}$, while for the
150\,Myr old M35 $\log_{10}\Q < 4.6$. Outside the $2\,\text{days} <
P_{\text{tide}} < 10\,\text{days}$ range the dissipation was not well
constrained. The difference in tidal periods probed by the two studies is due to
the fact that much stronger tides are required to affect the orbit than the spin
of the primary star.

\citet{Penev2018} (P18 from now on) investigated the effect of tides on the spin
of exoplanet host stars. Thanks to their small mass, the tides planets produce
on their parent stars are much weaker than binary star tides. As a result,
studying exoplanet systems allows probing tidal dissipation at even shorter
orbital (and hence tidal) periods than this or any of the above studies. P18
found a highly frequency-dependent dissipation, $\Q\propto P_{tide}^{-3.1}$, for
tidal periods $0.5\,\mathrm{days} < P_{tide} < 2\,\mathrm{days}$.

It should be noted that, while in Fig.~\ref{fig:merged_constraints} we plot the
inferred tidal dissipation as a function of tidal frequency, that need not
necessarily be the explanation of all differences observed. Each article
analyzes a different collection of binary star or exoplanet systems with
different selection biases involved. Given the wide variety of possible tidal
dissipation physics (see Sec. \ref{sec:introduction}), differences between measured
dissipation in different studies may be caused by any systematic difference
between the populations investigated, or even by trends  within a population.
For example, the dominant tidal period in each system is related to the orbital
period, but eccentricity is also tightly correlated with orbital period. As a
result, the binaries driving the tightly constrained $\Q$ at tidal periods
$>15$\,days are the longest period binaries in our sample, which also have
systematically higher eccentricities than the binaries dominating the
constraints in PS22, and even more compared to the P18 systems which are
virtually all circular. In turn, the orbital eccentricity determines the
combination of tidal waves each component of the binary is subjected to.

\subsection{Comarison with Studies By Other Groups}
\label{sec:compare_others}

Our group is far from the only one attempting to infer how efficiently tidal
perturbations are dissipated by studying signatures of tidal dissipation in
observed populations of objects. An exhaustive review of all such efforts in the
literature is beyond the scope of this article, but we highlight some of the
most recent efforts here to allow comparison to our results.

One approach is to directly compare the predictions of a particular tidal
dissipation model to observation. This was recently done by \citet{Barker_2022}
who compared the predictions of inertial wave enhanced dissipation to observed
tidal circularization, and found good agreement. In that model, the dissipation
is strongly enhanced if the tidal period is more than half the spin period of the
star. On the main sequence, that model predicts $\Q\sim10^7
(P_\star/10\,\mathrm{days})^2$ for Sun-like stars, where $P_\star$ is the spin
period of the star. This corresponds to several times more efficient dissipation
than what we find at the longest orbital periods in our sample, and several times
less efficient than the dissipation PS22 find for main sequence binaries
(assuming synchronous rotation).

\citet{Meibom_Mathieu_05} and \citet{Milliman_et_al_14} used observed orbits of
spectroscopic binary stars in open clusters to claim a detection of continued
circularization during the main sequence for Sun-like stars that could be
explained by $\Q\sim\mathrm{few}\times10^5$ for tidal periods of several days.
This overlaps reasonably well with the PP22 results and lies within a factor of two
of the lower limit this study finds for such orbital periods.

\citet{Zanazzi_2022} recently combined the W19 eccentricities with
similar constraints from TESS \citep{Justesen_2021} to argue that the
period-eccentricity distribution of binary stars should be split into two
components: a cold core with very close to zero eccentricity out to orbital
periods as large as 10 days, and an envelope of systems which is only
circularized out to periods of 3 or 4 days, with a broad distribution of
eccentrities after that. The cold core requires tidal  dissipation similar to
what PS22 and \citet{Meibom_Mathieu_05} find, while the envelope needs much less
efficient dissipation, perhaps arguing for a dissipation mechanism that only
operates on some systems and not on others based on initial conditions.

\citet{Hamer2019} argue that the galactic velocity dispersion of Hot Jupiter
host stars implies they are on average younger than field stars, interpreting
this as evidence for tidal inspiral of Hot Jupiters over the main sequence
lifetime of Sun-like stars. This in turn requires $\Q<10^7$, in agreement with
all constraints mentioned above.

\subsection{Caveats and Assumptions}
\label{sec:caveats}

The analysis presented here makes a number of assumptions. Below we discuss
some of those assumptions.

The most central assumption we make is that each of the binaries in our input
sample contains no other dynamically relevant objects and that tides are the
only thing causing the orbit to evolve or stellar spin to deviate from that of
an isolated star. While triple or higher multiplicity stellar systems are not
uncommon, the W19 analysis used 4 years of ultra-high precision photometric
measurements which result in high sensitivity to eclipse timing variations that
would be produced by the presence of additional objects. Indeed, W19 encounter
such objects and discard them from their sample. Nonetheless, some higher
multiplicity systems likely still remain in the sample, though the configurations
of those systems will be such that the extra objects likely do not affect the
dynamics too strongly.

A critical component of our model affecting the predicted stellar spins is the
loss of angular momentum to stellar winds. The model we use correctly captures
the major features for stars spinning no slower than the Sun, but for slow
rotators, \citep{vanSaders_et_al_16, Metcalfe_Egeland_19} argue spin-down stalls.
Luckily, the slowest spinning star in our sample has a spin period of 21 days,
which is shorter than the period at which spin-down may stall.

Different tidal dissipation theories predict effective $\Q$ that depends on a
whole host of parameters (mass, age, metallicity, and spin of the star;
amplitude of the tidal distortion; etc.). Here we allowed only for smooth
dependence on the frequency with which each tidal wave propagates on the surface
of the star, but all the other parameters change from one system to the next.
Some parameters are even correlated with tidal frequency (e.g. tidal amplitude
    gets smaller at longer orbital periods which also correspond to longer tidal
periods). Not accounting for such dependencies could skew, or even mask
completely, the frequency dependence. The fact that the constraints we obtain
for individual systems are consistent with each other is an encouraging sign
that such dependencies are not dominant; however, that possibility cannot be
ruled out entirely. Investigating dependencies of $\Q$ on multiple parameters
would require a significantly larger sample of systems than the 70 used here.

We assumed the tidal dissipation to be present only in the convective zone of
the star. Although this is what equilibrium tide models predict for stars with
convective envelopes, some dynamical tide models indicate that the dissipation
will be dominated by interactions with g-mode waves inside the star's core. For
the same exact $\Q$, such models will predict significantly different surface
spin for stars, because tides will deposit angular momentum in the core, which
will be communicated to the envelope on some coupling timescale. Thus, somewhat
more efficient dissipation will be required to explain the observed spins if
tides couple to the core.  Exploring the implications of tides coupling to the
radiative core instead of the convective envelope is a separate project of
comparable complexity to the one presented in this article. As such we consider
it outside the scope at this moment and may pursue it in the future.

The treatment of angular momentum redistribution within stars in POET is quite
simplistic. The core and the envelope are assumed to converge to synchronous
rotation on a fixed timescale, and each zone is assumed to follow solid body
rotation. While this is probably a reasonable assumption for the core, we know
this picture is not correct for the convective envelope. For example, in the
Sun, different parts of the convective zone spin differently, and are not even
organized in spherical shells (see for example \citep{Miesch2005} for more
details). Since we compare predicted to observed spins to infer the
tidal dissipation effeciency, our results could be affected if important physics
is not captured by the simplistic treatment used.

%% file: conclusion.tex
\section{Conclusion}\label{sec:conclusion}

By combining catalogues from the literature of the physical properties and
primary star spins of a collection of eclipsing binaries, we were able to infer
how efficiently tidal perturbations dissipate energy. We find tight constraints
for a frequency dependent modified tidal quality factor $Q_{\ast}^{\prime}$ for
tidal frequencies between 15 and 50 days, with $7.8 < \log_{10}\Q < 8.1$ at the
longest periods, with indications of decreasing by approximately 0.5\,dex
(dissipation efficiency enhanced by a factor of 3) for periods of 15 days. Below
15 days the precision with which the dissipation is measured is dramatically
reduced: $5.7 < \log_{10}\Q < 7.5$. The decreased precision is because most
systems sensitive to such tidal periods are also in close enough orbits to have
their spin synchronized with the orbit, reducing our ability to place a lower
limit on $\Q$ (upper limit on the dissipation).

The constraints obtained in this article expand on previous constraints found by our
group, which has embarked on a long-term effort to explore signatures of tidal
dissipation in low mass stars in multiple settings, including tidal
circularization in binary stars, the effects of tides on the spin exoplanet
hosts and binary star systems, and others. Comparison between all constraints in
this series to date is presented in Fig.~\ref{fig:merged_constraints}.
Interpreting the variations of $\Q$ from one system to the next as a frequency
dependence, we find a consistent picture, with constraints from different
contexts overlapping with each other for tidal frequencies probed by multiple
analyses.

By combining multiple measurements of $\Q$ we plan to explore its dependencies
on various parameters, helping to narrow down the dominant physical mechanisms
for the dissipation in various regimes. Such a multi-faceted approach is made
possible by the recent and on-going explosion of observations (soon to be)
produced by missions like \textit{Kepler}, TESS, Gaia, PLATO etc.

In service of this long term goal, our analysis is designed to account for
observational and model uncertainties as fully as possible, taking special care
to avoid conclusions not strongly supported by observations. We also avoid
assuming any particular tidal dissipation model, relying on a general parametric
prescription instead that is able to accommodate to a reasonable degree all the
proposed models.

%% file: data_availability.tex
\section*{Data Availability}
\label{sec:data_availability}

All tables of input parameters presented in this article are derivative of
published tables already available in machine readable form through the VizieR
online service. They are reproducible by applying simple filtering to those
tables, and thus do not warrant creating and publishing independent copies.

We have created a zenodo archive \citep{zenodo_results} to accompany this
article, which provides machine readable tables of the following:

\begin{itemize}
    \item The generated MCMC samples for each binary in the original HDF5 format
        produced by the emcee package \citep{Foreman_Mackey_2013} (see
        \url{https://emcee.readthedocs.io/en/stable/user/backends/}).
    \item The 2.3\%, 15.9\%, 84.1\%, and 97.7\% quantiles of $\log_{10}\Q$ for
        the given binary as a function of tidal period for each binary  (i.e.
            the coordinates defining the quantile curves in
            Fig.~\ref{fig:ind_const_1} and \ref{fig:ind_const_2}).
    \item The burn-in period for each quantile vs. tidal period (i.e. the
            coordinates defining the burn-in curves in Fig.~\ref{fig:burnin_1}
            and \ref{fig:burnin_2}).
    \item The estimated standard deviation of the fraction of samples below each
        of the 4 target quantiles for each binary (i.e. the coordinates defining
            the curves in Fig.~\ref{fig:cdfstd_1} and \ref{fig:cdfstd_2}).
    \item The 2.3\%, 15.9\%, 84.1\%, and 97.7\% percentiles of the combined
        $\log_{10}\Q$ constraint from all binaries vs tidal period (i.e. the
        coordinates of the quantile curves in Fig.~\ref{fig:combined}).
    \item The combined constraint 2.3\%, 15.9\%, 84.1\%, and 97.7\% percentiles
        and the 2.3\% and 97.7\% percentiles of individual constraints at the
        tidal periods where the difference between the latter is smallest (i.e.
            the coordinates defining the curves and points in
            Fig.~\ref{fig:combined_vs_individual}).
\end{itemize}

The version of POET used for calculating the orbital evolution is available
through zenodo archive \citet{zenodo_poet}.

%% file: q_tweak.tex
\section{Modified Tidal Dissipation Prescription}
\label{sec:q_tweak}

If $\alpha<0$, Eq. \ref{eq:q_formalism} has a discontinuity at
$\Omega_{m,m^{\prime}} = 0$. This makes it possible for tides to hold the system
in a spin orbit lock, keeping $\Omega_{m,m^{\prime}} = 0$ for an extended period
of time. \texttt{POET} handles this discontinuity by eliminating the stellar
spin as a variable and checking at each time step that the lock is maintained.
For $\alpha>0$, there is no discontinuity. However, even in this case it often
still happens that the system is held close to a lock for a long time, which
forces the differential equation solver to use tiny time steps in order to
maintain numerical accuracy. This in turn makes such cases very computationally
expensive to simulate. In order to avoid this situation, we slightly modify Eq.
\ref{eq:q_formalism} if $\alpha>0$ to force a small, but finite dissipation as
$\Omega_{m,m^{\prime}} \to 0$:

\begin{equation}
    \label{eq:q_tweak}
    \Delta_{m,m^{\prime}}
    =
    \Delta_0
    \begin{cases}
        {\left(\dfrac{\Omega_{\text{min}}}{\Omega_{\text{break}}}\right)}^{\alpha}
        & \text{if } \Omega_{m,m^{\prime}} < \Omega_{\text{min}} \\
        {\left(\dfrac{\Omega_{m,m^{\prime}}}{\Omega_{\text{break}}}\right)}^{\alpha}
        & \text{if }  \Omega_{\text{min}} < \Omega_{m,m^{\prime}} <
        \Omega_{\text{break}} \\
        1 & \text{if } \Omega_{m,m^{\prime}} > \Omega_{\text{break}}
    \end{cases}
\end{equation}

We pick $\Omega_{min}=\frac{2\pi}{50\,d}$ in Eq. \ref{eq:q_tweak}. This is small
enough to not appreciably change the evolution, while at the same time
eliminating most of the computational challenges.